%% file: Final-v3-v4.tex
\documentclass[a4paper,11pt]{article}
\usepackage{a4wide,amsmath,amssymb,bbm}
\usepackage[english]{babel}

\pdfoutput=1

\makeatletter

\@addtoreset{equation}{section} \makeatother

\usepackage{titletoc}
\dottedcontents{section}[6mm]{\smallskip }{6mm}{0pc}
\dottedcontents{subsection}[8mm]{\itshape\small }{6mm}{0pc}
\dottedcontents{subsubsection}[16mm]{\itshape\small }{10mm}{0pc}
\usepackage[nosort]{cite}

\usepackage[parsep]{collref}

\usepackage{graphicx}
\usepackage{feynmp}
\DeclareGraphicsExtensions{.pdf}  % include pdf files by default

\def \bea  {\begin{eqnarray}}
\def \eea  {\end{eqnarray}}

\newcommand{\ket}[1]{|{#1}\rangle}
\newcommand{\nn}{\nonumber}

\usepackage{parskip}

\def\unit{\protect{{1 \kern-.28em {\rm l}}}}

\begin{document}
%% Feynman diagrams
\input{dgms.tex}

\hfill{Imperial-TP-LW-2015-02}

\vspace{50pt}

\begin{center}
{\LARGE{\bf The $AdS_n\times S^n\times T^{10-2n}$ BMN string at two loops}}

\vspace{50pt}

{\bf \large  Per Sundin$^a$} and {\bf\large Linus Wulff$\, ^b$}

\vspace{15pt}

{$^a$ \it\small Universit\'a di Milano-Bicocca and INFN Sezione di Milano-Bicocca,\\ Dipartimento de Fisica,
Piazza della Scienza 3, I-20126 Milano, Italy}\\
\vspace{5pt}
{$^b$ \it\small Blackett Laboratory, Imperial College, London SW7 2AZ, U.K.}\\

\vspace{90pt}

{\bf Abstract}

\end{center}
\noindent
We calculate the two-loop correction to the dispersion relation for worldsheet modes of the BMN string in $AdS_n\times S^n\times T^{10-2n}$ for $n=2,3,5$. For the massive modes the result agrees with the exact dispersion relation derived from symmetry considerations with no correction to the interpolating function $h$. For the massless modes in $AdS_3\times S^3\times T^4$ however our result does not match what one expects from the corresponding symmetry based analysis. We also derive the S-matrix for massless modes up to the one-loop order. The scattering phase is given by the massless limit of the Hern\'andez-L\'opez phase. In addition we compute a certain massless S-matrix element at two loops and show that it vanishes suggesting that the two-loop phase in the massless sector is zero.

\pagebreak 
\tableofcontents
\setcounter{page}{1}

%%%%%%%%%%%%%%%%%%%%%%%%%%%%%%%%%%%%%%%%%%%%%%%%%%%%%%%%%%%%%%%%%%%%%%%%

\section{Introduction}
The point-like BMN string solution plays a special role in the AdS/CFT-correspondence. Expanding around this solution gives a systematic way to compute corrections to the anomalous dimensions of certain long single-trace operators in the CFT from quantum corrections to the energy of the string \cite{Berenstein:2002jq}. From the point of view of the string worldsheet theory one can fix a light-cone gauge adapted to the BMN geodesic to obtain a free theory of 8 massive (and massless in general) bosons and fermions plus an infinite number of interaction terms suppressed by inverse powers of the string tension. One can then carry out perturbative calculations in this (non-relativistic) 2d field theory. Here we will be interested in examples for which the worldsheet theory is known to be classically integrable (and believed to be quantum integrable). In many such examples one can guess the exact S-matrix and dispersion relations for the 2d theory from symmetries with a few extra assumptions. Direct perturbative calculations then give a valuable test of these arguments and the assumptions made.

From the point of view of perturbative calculations it turns out that the simplest examples are strings in $AdS_n\times S^n\times T^{10-2n}$ for $n=2,3,5$.\footnote{The corresponding supergravity backgrounds preserve 8, 16 and 32 supersymmetries respectively. For the discussion of classical integrability see the original papers \cite{Bena:2003wd,Babichenko:2009dk,Sundin:2012gc,Sorokin:2011rr,Cagnazzo:2011at} or, for a general treatment for strings on symmetric spaces, see \cite{Wulff:2015mwa}.} The reason for this is that in these models one has only even order interaction vertices in the BMN expansion. In particular there are no cubic interactions which leads to a big reduction in the number of possible Feynman diagrams compared to other cases such as $AdS_4\times\mathbbm{CP}^3$ and $AdS_3\times S^3\times S^3 \times S^1$ \cite{Zarembo:2009au, Rughoonauth:2012qd}. 

For a long time progress in going beyond tree-level was hampered by the fact that it was not clear how to deal with the divergences that show up and how to regularize in a way that preserves the symmetries expected of the answer. Therefore calculations were done either in the so-called Near-Flat-Space limit \cite{Maldacena:2006rv,Klose:2007wq,Klose:2007rz,Puletti:2007hq,Murugan:2012mf,Sundin:2013ypa}, using generalized unitarity \cite{Bianchi:2013nra,Engelund:2013fja,Bianchi:2014rfa} or by computing only quantities that were explicitly finite \cite{Rughoonauth:2012qd,Sundin:2012gc,Abbott:2013kka,Sundin:2014sfa}. Thus one could side-step the issue of regularization. Recently however this hurdle was overcome in \cite{Roiban:2014cia} where it was shown how to compute the one-loop correction to the dispersion relation and S-matrix by correctly treating the divergences as wave function renormalization and using a scheme for reducing the integrals that appeared to a smaller subset which could be easily computed. The same ideas were then applied to the $AdS_3\times S^3\times T^4$ string with a mix of NSNS and RR flux and to the calculation of the two-loop dispersion relation for both massive and massless modes in the Near-Flat-Space limit in \cite{Sundin:2014ema}.

Here we will extend these techniques to compute for the first time\footnote{The two-loop partition function for folded strings in $AdS_5\times S^5$, related to the cusp anomalous dimension on the gauge theory side, has been computed in \cite{Roiban:2007jf,Roiban:2007dq,Roiban:2007ju,Giombi:2009gd,Giombi:2010fa,Giombi:2010zi,Iwashita:2011ha}.
} the full BMN two-loop dispersion relation. The result takes the following form
\begin{align}
\label{eq:bmn-dispersion-relation}
AdS_5\times S^5:&\,\,\,\varepsilon^2=1+p_1^2-\frac{p_1^4}{12g^2}+\mathcal O(g^{-3})\,,\\
AdS_3\times S^3\times T^4:&\,\,\,
\varepsilon^2_{(m=\hat q)}=\hat q^2+p_1^2-\hat q^2\frac{(p_1\mp q)^4}{12g^2}+\mathcal O(g^{-3})\,,\quad 
\varepsilon^2_{(m=0)}=p_1^2-\hat q^2\frac{p_1^4}{2\pi^2g^2}+\mathcal O(g^{-3})\,,\nonumber\\
AdS_2\times S^2\times T^6:&\,\,\,
\varepsilon^2_{(m=1)}=1+p_1^2-\frac{p_1^4}{48g^2}+\mathcal O(g^{-3})\,,\quad 
\varepsilon^2_{(m=0)}=p_1^2-\frac{p_1^4}{4\pi^2g^2}+\mathcal O(g^{-3})\,.\nonumber
\end{align}
For $AdS_5\times S^5$ and the massive modes in $AdS_3\times S^3\times T^4$ the result agrees with the expansion of the proposed exact dispersion relation suggested in \cite{Hoare:2013lja,Lloyd:2014bsa} which takes the form
\begin{equation}
\varepsilon^2_\pm=(1\pm qg\mathrm p)^2+4\hat q^2h^2\sin^2\frac{\mathrm p}{2}\,,
\label{eq:disp-rel}
%=
%\hat q^2+p_1^2
%-\hat q^2\frac{(p_1\mp q)^4}{12g^2}
%\pm 2qp(-f/g^2)
%-2q^2(-f/g^2)
%\mp 2pq^3(-2f/g^2)
%+q^2(p^2+q^2)(-2f/g^2)
%+\mathcal O(g^{-3})
%
%h=g+f/g+O(g^{-2})
\end{equation}
provided that we take $h=g+O(g^{-2})$, i.e. the interpolating function receives no corrections up to two loops (strictly speaking this is needed only when $q\neq0$). Here $0\leq q\leq1$ measures the amount of NSNS flux of the $AdS_3\times S^3\times T^4$ background and $\hat q=\sqrt{1-q^2}$ measures the amount of RR flux. The $AdS_5\times S^5$ dispersion relation is obtained by setting $q=0$. To compare this dispersion relation with the one we calculate in the BMN limit one needs to rescale the spin-chain momentum as $\mathrm p=(p_1\mp q)/h$ (the shift by $q$ is needed in the comparison because we have defined the massive modes such that their quadratic action is Lorentz-invariant, see footnote \ref{foot:q-redef}). For $AdS_2\times S^2\times T^6$ the symmetries are not enough to completely fix the form of the dispersion relation \cite{Hoare:2014kma} but nevertheless we find the same two-loop correction as for the other cases but with $g$ replaced by $2g$ (the same was observed at one loop in \cite{Roiban:2014cia}).

For the massless modes in $AdS_3\times S^3\times T^4$ the exact dispersion relation was suggested, based on a similar symmetry argument that gave the massive mode dispersion relation, to have the same form as in (\ref{eq:disp-rel}) except with the $1$ in the first term replaced by $0$ \cite{Lloyd:2014bsa}. Here however the worldsheet calculation is not in agreement (note that the BMN rescaling is $\mathrm p=p_1/h$ in this case without the $q$-shift). This was already noted in the Near-Flat-Space limit in \cite{Sundin:2014ema}.\footnote{As remarked there this means that the discrepancy is not due to one calculation being done in type IIA and the other in type IIB as the string action is the same for the two cases in the NFS-limit. In fact we have explicitly checked that the correction to the dispersion relation is the same in the type IIB case.} Though we find the same form of the dispersion relation to two loops the coefficient of the two-loop correction differs by a factor of $6/\pi^2$. The $1/\pi^2$-factor can be traced to the types of integrals that contribute in this case, they have one massless and two massive modes running in the loops, and are thus quite different from the integrals with three massive modes which contribute to the massive dispersion relation. As speculated in \cite{Sundin:2014ema} this mismatch could be due to a misidentification of the asymptotic states in the two approaches or due to unexpected quantum corrections to the central charges. Unfortunately we will not be able to resolve it here. 

In order to  gain further insight into the role of the massless modes we also probe the worldsheet S-matrix in the massless sector of the theory of $AdS_3\times S^3\times T^4$. At tree-level we find that there is no phase contribution and at one loop the dressing phase is simply given by the massless limit of the well known Hern\'andez-L\'opez phase \cite{Beisert:2006ib}(up to an IR-divergent piece arising from the limit taken). We furthermore find, to the extent the type IIA and type IIB comparison is valid,  that our results are consistent with the symmetry based analysis of \cite{Borsato:2014hja}. 

We then push the analysis to the two-loop level and compute a forward type scattering element of two massless bosons. Since we are using the full BMN-string, where the relevant vertices are fourth, sixth and eight order in transverse fields, a large class of distinct Feynman diagrams contribute. However, once we go on-shell, and focus on the kinematic regime where the sign of the transverse momenta of the scattered particles is opposite, we find that the contribution from each topology vanishes separately. That is, each integral is multiplied with high enough powers of the external momenta to vanish once we go on-shell. Thus the entire two-loop part of the massless S-matrix, for this specific scattering element, is zero. This indicates that there should be no contribution from the phase at this order in perturbation theory. 

The outline of the paper is as follows: In section \ref{sec:GS} and \ref{sec:BMN-expansion} we write down the string action and perform the BMN expansion. In section \ref{sec:disp-relation} we evaluate the two-loop dispersion relation and explain in detail how to regularize the divergent integrals that appear in the computation. Finally, in section \ref{sec:massless-S} we analyze the massless S-matrix up to the two-loop order and compare with the exact solution, to the extent it is known. We end the paper with a summary and outlook. Some details of the dispersion relation calculation are deferred to appendices.

\section{Green-Schwarz string action}
\label{sec:GS}
The Green-Schwarz superstring action can be expanded order by order in fermions as\footnote{The sign of the action is due to using opposite conventions for the 2d and 10d metric.}
\begin{equation}
S=g\int d^2\xi\,(\mathcal L^{(0)}+\mathcal L^{(2)}+\ldots)\,,
\end{equation}
where $g$ is the string tension. In $AdS_5\times S^5$ this expansion is known to all orders due to the background being maximally supersymmetric \cite{Metsaev:1998it}. In a general type II supergravity background however, the expansion is only known explicitly up to quartic order \cite{Wulff:2013kga}. This is the action we will use for the string in $AdS_3\times S^3\times T^4$ and $AdS_2\times S^2\times T^6$. Its form is as follows. The purely bosonic terms in the Lagrangian are given by
\begin{equation}
\mathcal L^{(0)}=\frac12\gamma^{ij}e_i{}^ae_j{}^b\eta_{ab}+\frac12\varepsilon^{ij}B^{(0)}_{ij}\,,\qquad(\gamma^{ij}=\sqrt{-h}h^{ij})
\label{eq:L0}
\end{equation}
where we denote the purely bosonic vielbeins by $e^a$ $(a=0,\ldots,9)$ and $B_{ij}^{(0)}=e_i{}^ae_j{}^bB^{(0)}_{ab}$ is the lowest component in the $\Theta$-expansion of the NSNS two-form superfield $B$. The terms quadratic in fermions take the form\footnote{These expressions refer to type IIA. To get the type IIB expressions on should replace the $32$-component Majorana spinor $\Theta^\alpha$ by a doublet of $16$-component Majorana-Weyl spinors $\Theta^{\alpha' i}$ $i=1,2$ and the $32\times32$ gamma-matrices $\Gamma^a$ by $16\times16$ ones
\begin{equation}
\Gamma_a\rightarrow\gamma_a\,,\qquad\Gamma_{11}\rightarrow\sigma^3\qquad(\mbox{except:}\quad\Gamma_{11}T\rightarrow-\sigma^3T)\,.\nonumber
\end{equation}
Finally, instead of the $\mathcal S$ defined in (\ref{eq:SbA}) one should use the expression appropriate to type IIB
\begin{equation}
\label{eq:SbB}
\mathcal S=-e^\phi\big(i\sigma^2\gamma^a F^{(1)}_a+\frac{1}{3!}\sigma^1\gamma^{abc}F^{(3)}_{abc}+\frac{1}{2\cdot5!}i\sigma^2\gamma^{abcde} F^{(5)}_{abcde}\big)\,.
\end{equation}
For more details and definitions of the gamma-matrices see \cite{Wulff:2013kga}.
}
\begin{equation}
\mathcal L^{(2)}=\frac{i}{2}e_i{}^a\,\Theta\Gamma_aK^{ij}\mathcal D_j\Theta\,,\qquad K^{ij}=\gamma^{ij}-\varepsilon^{ij}\Gamma_{11}\,,
\end{equation}
where the Killing spinor derivative is defined as
\begin{equation}
\label{eq:DbA}
\mathcal D\Theta=
\big(d-\frac{1}{4}\omega^{ab}\Gamma_{ab}+\frac{1}{8}e^aG_a\big)\Theta\,,\qquad G_a=H_{abc}\,\Gamma^{bc}\Gamma_{11}+\mathcal S\Gamma_a\,,
\end{equation}
$\omega^{ab}$ is the spin connection, $H=dB$ is (the bosonic part of) the NSNS three-form field strength and the type IIA RR fields enter the action through the bispinor
\begin{equation}
\label{eq:SbA}
\mathcal S=e^\phi\big(\frac12F^{(2)}_{ab}\Gamma^{ab}\Gamma_{11}+\frac{1}{4!}F^{(4)}_{abcd}\Gamma^{abcd}\big)\,.
\end{equation}
Finally the quartic terms in the Lagrangian take the form
\begin{align}
\mathcal L^{(4)}=&
-\frac{1}{8}\Theta\Gamma^a\mathcal D_i\Theta\,\Theta\Gamma_aK^{ij}\mathcal D_j\Theta
+\frac{i}{24}e_i{}^a\,\Theta\Gamma_aK^{ij}\mathcal M\mathcal D_j\Theta
+\frac{i}{192}e_i{}^ae_j{}^b\,\Theta\Gamma_aK^{ij}(M+\tilde M)\mathcal S\Gamma_b\Theta
\nonumber\\
&{}
+\frac{1}{192}e_i{}^ce_j{}^d\,\Theta\Gamma_c{}^{ab}K^{ij}\Theta\,(3\Theta\Gamma_dU_{ab}\Theta-2\Theta\Gamma_aU_{bd}\Theta)
\nonumber\\
&{}
-\frac{1}{192}e_i{}^ce_j{}^d\,\Theta\Gamma_c{}^{ab}\Gamma_{11}K^{ij}\Theta\,(3\Theta\Gamma_d\Gamma_{11}U_{ab}\Theta+2\Theta\Gamma_a\Gamma_{11}U_{bd}\Theta)\,.
\label{eq:L4}
\end{align}
Where we have defined two matrices which are quadratic in fermions
\begin{align}
\mathcal M^\alpha{}_\beta=&{}
M^\alpha{}_\beta
+\tilde M^\alpha{}_\beta
+\frac{i}{8}(G_a\Theta)^\alpha\,(\Theta\Gamma^a)_\beta
-\frac{i}{32}(\Gamma^{ab}\Theta)^\alpha\,(\Theta\Gamma_aG_b)_\beta
-\frac{i}{32}(\Gamma^{ab}\Theta)^\alpha\,(C\Gamma_aG_b\Theta)_\beta
\nonumber\\
M^\alpha{}_\beta=&{}
\frac12\Theta T\Theta\,\delta^\alpha_\beta
-\frac12\Theta\Gamma_{11}T\Theta\,(\Gamma_{11})^\alpha{}_\beta
+\Theta^\alpha\, (CT\Theta)_\beta
+(\Gamma^aT\Theta)^\alpha\,(\Theta\Gamma_a)_\beta
\label{eq:M}
\end{align}
while $\tilde M=\Gamma_{11}M\Gamma_{11}$. In addition two new matrices constructed from the background fields contracted with gamma-matrices appear at this order
\begin{align}
T=\frac{i}{2}\nabla_a\phi\,\Gamma^a+\frac{i}{24}H_{abc}\,\Gamma^{abc}\Gamma_{11}+\frac{i}{16}\Gamma_a\mathcal S\Gamma^a\,,\quad
U_{ab}={}\frac{1}{4}\nabla_{[a}G_{b]}+\frac{1}{32}G_{[a}G_{b]}-\frac{1}{4}R_{ab}{}^{cd}\,\Gamma_{cd}\,.
\label{eq:T}
\end{align}
The first appears in the dilatino equation and the second in the integrability condition for the Killing spinor equation. Due to this fact $T\Theta$ and $U_{ab}\Theta$ will be proportional to the non-supersymmetric (non-coset) fermions in symmetric space backgrounds such as the ones we are interested in.

\section{BMN expansion in $AdS_n\times S^n\times T^{10-2n}$}\label{sec:BMN-expansion}
The string action simplifies in the cases we are considering since all background fields are constant. The backgrounds we consider are supported by the following combinations of fluxes (see \cite{Wulff:2014kja} for conventions)
\begin{align}
AdS_5\times S^5:&\quad F^{(5)}=4e^{-\phi}(\Omega_{AdS_5}+\Omega_{S^5})\,,\\
AdS_3\times S^3\times T^4:&\quad F^{(4)}=2\hat qe^{-\phi}dx^9(\Omega_{AdS_3}+\Omega_{S^3})\,, \quad H=2q(\Omega_{AdS_3}+\Omega_{S^3})\,,\\
AdS_2\times S^2\times T^6:&\quad F^{(4)}=2e^{-\phi}([dx^7dx^6-dx^9dx^8]\Omega_{AdS_2}+[dx^8dx^6+dx^9dx^7]\Omega_{S^2})\,,
\end{align}
where $AdS_3\times S^3\times T^4$ is supported by a combination of NSNS- and RR-flux parameterized by $\hat q, q$ satisfying 
$$
\hat q^2 + q^2=1\,.
$$ 
Note that for $AdS_3\times S^3\times T^4$ and $AdS_2\times S^2\times T^6$ we are taking the type IIA solutions but we could of course also have used the IIB solutions obtained by T-duality along a torus direction. The $AdS$ and $S$ radius are both set to one in these conventions. From (\ref{eq:SbA}), (\ref{eq:SbB}) and (\ref{eq:T}) we find
\begin{align}
AdS_5\times S^5:&\quad \mathcal S=-4\varepsilon\gamma^{01234}\,,\quad T=0\,,\\
AdS_3\times S^3\times T^4:&\quad \mathcal S=-4\hat q\mathcal P_{16}\Gamma^{0129}\,,\quad T=-\frac{i}{2}\mathcal P_{16}\Gamma^{012}(\hat q\Gamma^9+2q\Gamma_{11})\,,\\
AdS_2\times S^2\times T^6:&\quad \mathcal S=-4\mathcal P_8\Gamma^{0167}\,,\quad T=-\frac{i}{2}\mathcal P_8\Gamma^{0167}\,,
\end{align}
where we have defined the following projection operators
\begin{equation}
\mathcal P_{16}=\frac12(1+\Gamma^{012345})\,,\qquad\mathcal P_8=\frac12(1+\Gamma^{6789})\frac12(1+\Gamma^{012378})\,,
\end{equation}
the index indicating the dimension of the space they project on, i.e. the number of supersymmetries preserved in each case.

The $AdS$-metric is taken to be
\begin{equation}
ds^2_{AdS_n}=-\left(\frac{1+\frac12|z_I|^2}{1-\frac12|z_I|^2}\right)^2dt^2+\frac{2|dz_I|^2}{(1-\frac12|z_I|^2)^2}\qquad I=1,\ldots,(n-1)/2\,,
\label{eq:AdSmetric}
\end{equation}
where the transverse coordinates are grouped together into two complex coordinates in $AdS_5$, one in $AdS_3$ and one real coordinate, $x_1=\sqrt2\,z$, in $AdS_2$. Similarly the $S$-metric is
\begin{equation}
ds^2_{S^n}=\left(\frac{1-\frac12|y_I|^2}{1+\frac12|y_I|^2}\right)^2d\varphi^2+\frac{2|dy_I|^2}{(1+\frac12|y_I|^2)^2}\qquad I=1,\ldots,(n-1)/2\,.
\label{eq:Smetric}
\end{equation}
In these coordinates the NSNS two-form appearing in eq. (\ref{eq:L0}), which is only non-zero for $AdS_3\times S^3\times T^4$, takes the form
\bea
B^{(0)}=-iq\frac{zd\bar z-\bar zdz}{(1-\frac12|z|^2)^2}dt+iq\frac{yd\bar y-\bar ydy}{(1+\frac12|y|^2)^2}d\varphi\,.
%H=dB=-2iq\frac{1+\frac{1}{2}|z|^2}{(1-\frac{1}{2}|z|^2)^3}dzd\bar zdt
%+2iq\frac{1-\frac{1}{2}|y|^2}{(1+\frac{1}{2}|z|^2)^3}dyd\bar yd\varphi\,.
\eea 

Plugging the form of the background fields into the action the next step is to expand around the BMN-solution given by $x^+=\frac12(t+\varphi)=\tau$ \cite{Berenstein:2002jq}. At the same time we fix the light-cone gauge and corresponding kappa symmetry gauge
\begin{equation}
\label{eq:gf}
x^+=\tau\,,\qquad\Gamma^+\Theta=0\,.
\end{equation}
The Virasoro constraints are then used to solve for $x^-$ in terms of the other fields. In this gauge the worldsheet metric defined in (\ref{eq:L0}) takes the form $\gamma=\eta+\hat\gamma$, where $\hat\gamma$ denotes higher order corrections to be determined from the conditions on the momentum conjugate to $x^-$
\begin{equation}
\label{eq:gf1}
p^+=-\frac12\frac{\partial\mathcal L}{\partial \dot x^-}=1\,,\qquad\frac{\partial\mathcal L}{\partial x^-{}'}=0\,.
\end{equation}

Rescaling all transverse coordinates with a factor $g^{-1/2}$ yields a perturbative expansion in the string tension\footnote{The fact that only even orders appear is a special feature of the $AdS\times S\times T$ backgrounds that make them particularly suited to a perturbative treatment.}
\bea \nn
\mathcal{L}= \mathcal{L}_2+\frac{1}{g}\mathcal{L}_4+\frac{1}{g^2}\mathcal{L}_6+\dots
\eea
where the subscript denotes the number of transverse coordinates in each term. The quadratic Lagrangian $\mathcal{L}_2$ takes the form\footnote{Here $\partial_\pm=\partial_0\pm\partial_1$. For $AdS_3\times S^3\times T^4$ with mixed flux we have performed a field redefinition of the massive modes $z\rightarrow e^{-iq\sigma}z$ and $y\rightarrow e^{iq\sigma}y$ and a similar redefinition for the fermions which puts the quadratic action into a Lorentz-invariant form.
\label{foot:q-redef}
}
\begin{align}
\mathcal{L}_2=&
|\partial z_I|^2
-m^2|z_I|^2
+|\partial y_I|^2
-m^2|y_I|^2
+|\partial u_{I'}|^2
+i\bar\chi_L^r\partial_-\chi_L^r
+i\bar\chi_R^r\partial_+\chi_R^r
-m\left(\bar\chi_L^r\chi_R^r+\bar\chi_R^r\chi_L^r\right)
\nonumber\\
&{}
+i\bar\chi_L^{r'}\partial_-\chi_L^{r'}
+i\bar\chi_R^{r'}\partial_+\chi_R^{r'}\,,
\label{eq:quadratic-L}
\end{align}
where unprimed indices run over massive modes and primed indices run over massless modes and $m=1$ except for $AdS_3\times S^3\times T^4$ with mixed flux in which case $m=\hat q$. The spectrum can be summarized as follows
\[ \label{tab:charges}
\begin{tabular}{c|cc|cc}
 & \multicolumn{2}{c|}{\textbf{Coset/massive}} &  \multicolumn{2}{c}{\textbf{Non-coset/massless}}\tabularnewline
\hline \nn
 & Bosons & Fermions & Bosons & Fermions \\
\hline
& & & &\\[-5pt]
$AdS_5\times S^5$ & $z_1,z_2,\,y_1,y_2$ & $\chi_{L,R}^{1,2,3,4}$ & - & -\\[5pt]
$AdS_3\times S^3\times T^4$ & $z,\,y$ & $\chi^{1,2}_{L,R}$ & $u_1,u_2$ & $\chi^{3,4}_{L,R}$ \\[5pt]
$AdS_2\times S^2\times T^6$ & $x_1,\,x_2$ & $\chi^1_{L,R}$ & $u_1,u_2,u_3$ & $\chi^{2,3,4}_{L,R}$\\[5pt]
\hline
\end{tabular}
\]
\vspace{0.2cm}\\
Here all coordinates are complex except $x_1=\sqrt2\,z$ and $x_2=\sqrt2\,y$, originating from the transverse directions of $AdS_2$ and $S^2$ respectively.

\section{Two-loop dispersion relation}\label{sec:disp-relation}
We now turn to the problem of determining the two-loop correction to the two-point function, i.e. the correction to the dispersion relation. There are three different topologies of Feynman diagrams that appear. The first, and by far the most complicated, are the sunset diagrams
\bea \label{diagram:sunset}
\parbox[top][0.8in][c]{1.5in}{\fmfreuse{sunset}}
\eea
which lead to the following loop integrals
\bea
\label{eq:sunset-integral}
I^{rstu}_{Mmm}(p)=\int\frac{d^2kd^2l}{(2\pi)^4}\frac{k_+^rk_-^sl_+^tl_-^u}{(k^2-M^2)(l^2-m^2)((p-k-l)^2-m^2)}\,,
\eea
where the masses of the virtual particles are either all the same, $M=m$, or two the same and one different $M\neq m$. Generically the integrals are (power counting) UV-divergent and sometimes IR-divergent when massless particles are involved and must be regularized. Our procedure for regularizing and computing the relevant integrals is described in the next section.

The second type of Feynman topologies are the four-vertex bubble-tadpoles
\bea
\label{diagram:8}
\parbox[top][0.8in][c]{1.5in}{\fmfreuse{doubblebubble}}
\eea
\\
which lead to a combination of a tadpole integral\footnote{Note that $T^{rs}_m(P)$ can be expressed in terms of (a sum of) $T^{rs}_m(0)$ by shifting the loop variable. In the following we will assume that this is done and simply write $T^{rs}_m$.}
\bea 
\label{eq:tadpole-integral}
T^{rs}_m(P)=\int\frac{d^2k}{(2\pi)^2}\frac{k_+^rk_-^s}{(k-P)^2-m^2}
\eea 
and a bubble integral
\bea 
\label{eq:bubble-integral}
B^{rs}_m(P)=\int\frac{d^2k}{(2\pi)^2}\frac{k_+^rk_-^s}{(k^2-m^2)((k-P)^2-m^2)}\,,
\eea 
where both the bubble and tadpole have $P=0$. 

Finally we have the double-tadpoles built out of a six-vertex
\bea
\parbox[top][0.8in][c]{1.5in}{\fmfreuse{doubletadpole}}
\eea
which lead to a product of two tadpole integrals (\ref{eq:tadpole-integral}) and are the simplest to evaluate (although a bit cumbersome since the sixth order Lagrangian contains many terms).

\subsection{Regularization procedure}
Our regularization scheme is similar to the one found to work at one loop in \cite{Roiban:2014cia}. It is based on reducing the integrals that appear, via algebraic identities on the integrand and shifts of the loop variable, to less divergent, or finite, integrals plus tadpole-type integrals. In this way it turns out to be possible to push all UV-divergences into tadpole-type integrals which can then be easily regularized. In $AdS_2\times S^2\times T^6$ and $AdS_3\times S^3\times T^4$ there are also massless modes present which lead to potential IR-divergences. These turn out to be simpler to deal with and we do this simply by introducing a small regulator mass $\mu$ for these modes which is sent to zero at the end. IR-divergences turn out to cancel within each class of diagrams independently unlike the UV-divergences. In \cite{Roiban:2014cia} it was explained how to reduce the bubble integrals so here we will focus on the sunset integrals in (\ref{eq:sunset-integral}).

The first step is to use the simple identity
\bea
\label{eq:triv-identity}
\frac{k_+k_-}{k^2-m^2}=\frac{k^2}{k^2-m^2}=1+\frac{m^2}{k^2-m^2}
\eea
on the integrand. This leads to a sunset integral with a lower degree of divergence plus an integral with one less propagator, which, by shifting the corresponding loop momentum $k\rightarrow p-k-l$, leads to a product of two one-loop tadpole integrals (\ref{eq:tadpole-integral}). By repeating this process the sunset integrals we have to compute are reduced to the following ones
\bea
\nn
I^{r0s0}\,,\quad I^{0r0s}\,,\quad I^{r00s}\,,\quad I^{0rs0}\,.
\eea 
Integrals with $r+s > 1$ are still (power-counting) UV-divergent of course. Note that it is enough to compute the first and third of these as the second and fourth differ only by replacing a $+$-index by a $-$-index and vice versa. To evaluate these it is useful to start by considering the one-loop bubble integrals defined in (\ref{eq:bubble-integral}). Following \cite{Roiban:2014cia} we use the algebraic identity
\begin{equation}
\label{eq:B00-rel}
\frac{1}{(k-P)^2-m^2}-\frac{1}{k^2-m^2}=\frac{P_+k_-+P_-k_+-P^2}{((k-P)^2-m^2)(k^2-m^2)}
\end{equation}
on the integrand to derive the recursion relation\footnote{Here we have used shifts of the loop variable in the tadpole integrals that appear and the fact that $T_m^{rs}=0$ for $r\neq s$ by Lorentz-invariance.}
\begin{equation}
\label{eq:bubble-id1}
B_m^{r0}(P)=P_+B_m^{r-1,0}(P)-m^2\frac{P_+}{P_-}B_m^{r-2,0}(P)
%+\frac{1}{P_-}(T^{r,0}(P)-P_+T^{r-1,0}(P)-T^{r,0}(0))
\qquad(r\geq2)
\end{equation}
together with
\begin{equation}
B_m^{10}(P)=\frac{P_+}{2}B_m^{00}(P)
\end{equation}
and the same for $B_m^{0r}$ with the order of the indices switched and $P_+\leftrightarrow P_-$. We can now use these relations inside the sunset integrals and we find
\begin{align}
I^{r0s0}_{Mmm}(p)=&
\int\frac{d^2k}{(2\pi)^2}\frac{k_+^r}{k^2-M^2}B_m^{s0}(p-k)
\nonumber\\
=&
p_+I^{r,0,s-1,0}_{Mmm}(p)
-I^{r+1,0,s-1,0}_{Mmm}(p)
-m^2\tilde I^{r,0,s-2,0}_{Mmm}(p)
%+\int\frac{d^2k}{(2\pi)^2}\frac{k_+^r}{k^2-M^2}\frac{1}{P_-}(T^{s-1,0}(P)-P_+T^{s-2,0}(P)-T^{s-1,0}(0))
\qquad(s\geq2)
\end{align}
and
\begin{equation}
I^{r010}_{Mmm}(p)=\frac{p_+}{2}I^{r000}_{Mmm}(p)-\frac12I^{r+1,0,0,0}_{Mmm}(p)
\end{equation}
where
\begin{equation}
\tilde I^{r0s0}_{Mmm}(p)=\int\frac{d^2k}{(2\pi)^2}\frac{p_+-k_+}{p_--k_-}\frac{k_+^r}{k^2-M^2}B_m^{s0}(p-k)\,.
\end{equation}
When $m=0$ the latter integral does not contribute, however when $m\neq0$ we need to compute it. To do this we use the fact that to find the dispersion relation we only need to compute the integrals on-shell\footnote{To compute for example the wave function renormalization at two loops one would need to evaluate the sunset integrals off-shell. This would require a more sophisticated approach to the regularization.} and for $m\neq0$ the contributing sunset integrals (\ref{eq:sunset-integral}) have
 $p^2=M^2$. This in turn implies the algebraic identity
\begin{equation}
-\frac{1}{p_--k_-}-\frac{k_+}{k^2-M^2}=p_-\frac{p_+-k_+}{(p_--k_-)(k^2-M^2)}
\end{equation}
from which it follows that
\begin{equation}
p_-\tilde I^{r0s0}_{Mmm}(p)
=
-I^{r+1,0,s,0}_{Mmm}(p)
+\sum_{n=0}^r\binom{r}{n}p_+^{r-n}I_{0mm}^{n+1,0,s,0}(0)
\end{equation}
where we have shifted $k\rightarrow p+k$ in the last term. All the integrals appearing in the sum are in fact zero by Lorentz invariance since they are evaluated at $p=0$. Putting this together we have the relations
\begin{align}
I^{r0s0}_{Mmm}(p)=&
p_+I^{r,0,s-1,0}_{Mmm}(p)
-I^{r+1,0,s-1,0}_{Mmm}(p)
+\frac{m^2}{p_-}I^{r+1,0,s-2,0}_{Mmm}(p)
\qquad(s\geq2)
\nonumber\\
I^{r010}_{Mmm}(p)=&\frac{p_+}{2}I^{r000}_{Mmm}(p)-\frac12I^{r+1,0,0,0}_{Mmm}(p)
\label{eq:Ir0s0}
\end{align}
and the same for $I^{0r0s}$ with $p_+\leftrightarrow p_-$. For general masses $m,M$ these recursion relations allow us to solve for integrals of the type $I^{r0s0}$ in terms of integrals of the type $I^{r000}$, which turns out to be enough for our purposes. However, in the case when all masses are equal, i.e. $M=m$, we have the extra symmetry $I^{rstu}=I^{turs}$ which allows us to solve completely for $I^{r0s0}$ in terms of $I^{0000}$ and we find
\begin{equation}
I^{r0s0}_{mmm}(p)=
\left\{
\begin{array}{cc}
p_+^{r+s}I^{0000}_{mmm}(p) &r,s\quad\mbox{even}\\[5pt]
\frac{(-1)^{r+s+1}}{3}p_+^{r+s}I^{0000}_{mmm}(p) &\mbox{otherwise}
\end{array}
\right.
\label{eq:Ir0s0-mmm}
\end{equation}
and the same expression with $p_-$ instead of $p_+$ for $I^{0r0s}$.

Repeating the same steps for the $I^{r00s}$ integrals we find
\begin{align}
I^{r00s}_{Mmm}(p)=&
p_-I^{r,0,0,s-1}_{Mmm}(p)
-I^{r,1,0,s-1}_{Mmm}(p)
+\frac{m^2}{p_+}I^{r,1,0,s-2}_{Mmm}(p)
-m^2\binom{r}{s-1}p_+^{r-s}I_{0mm}^{s-1,1,0,s-2}(0)
\qquad(s\geq2)
\nonumber\\
I^{r001}_{Mmm}(p)=&\frac{p_-}{2}I^{r000}_{Mmm}(p)-\frac12I^{r100}_{Mmm}(p)
\end{align}
and using the fact that (note that the last term vanishes unless $r\geq s+1$)
\begin{equation}
I^{r10s}(p)=M^2I^{r-1,0,0,s}(p)+(-1)^s\binom{r-1}{s}p_+^{r-s-1}T_m^{ss}T_m^{00}\,,
\end{equation}
where we have shifted the loop variable in the tadpole-like integral and used the Lorentz symmetry of the measure, this becomes
\begin{align}
I^{r00s}_{Mmm}(p)=&
p_-I^{r,0,0,s-1}_{Mmm}(p)
-M^2I^{r-1,0,0,s-1}_{Mmm}(p)
+m^2p_-I^{r-1,0,0,s-2}_{Mmm}(p)
\qquad(r\geq1\,,\quad s\geq2)
\nonumber\\
&{}
+(-1)^s\binom{r-1}{s-1}p_+^{r-s}T^{00}[T_m^{s-1,s-1}-m^2T_m^{s-2,s-2}]
\nonumber\\
I^{r001}_{Mmm}(p)=&\frac{p_-}{2}I^{r000}_{Mmm}(p)-\frac{M^2}{2}I^{r-1,0,0,0}_{Mmm}(p)-\frac{p_+^{r-1}}{2}[T_m^{00}]^2
%
%0=I^{2003}-I^{3002}
%=
%-p_+T^{00}(0)[T^{11}(0)-T^{00}(0)]
%
\label{eq:Ir00s}
\end{align}
and the same for $I^{0rs0}$ with $p_+\leftrightarrow p_-$. It turns out that for these expressions to be consistent we must use a scheme where all power-like divergences are set to zero so that for example
\begin{equation}
T_m^{rr}=m^{2r}T_m^{00}\,,
\label{eq:tadpole-rel}
\end{equation}
and the last term in the first expression drops out. This is of course what we often do, e.g. in dimensional regularization, and was needed also at one loop \cite{Roiban:2014cia}. For general masses these recursion relations turn out to be enough for our purposes while in the special case that all masses are equal we can again use the symmetry in the indices to solve for $I^{r00s}$ completely in terms of $I^{0000}$
\begin{equation}
I^{r00s}_{mmm}(p)=
\left\{
\begin{array}{cc}
p_+^rp_-^sI^{0000}_{mmm}(p) &r,s\quad\mbox{even}\\[5pt]
\frac{(-1)^{r+s+1}}{3}p_+^rp_-^sI^{0000}_{mmm}(p)-\frac{1-(-1)^{\mathrm{min}(r,s)}}{4m^2}p_+^rp_-^s[T_m^{00}]^2 &\mbox{otherwise}
\end{array}
\right.
\,.
\label{eq:Ir00s-mmm}
\end{equation}
Note the appearance of divergences when $\mathrm{min}(r,s)$ is odd, coming from the tadpole integral $T_m^{00}$, which were absent in (\ref{eq:Ir0s0-mmm}). Let us recall again that in this derivation of the identities (\ref{eq:Ir0s0}), (\ref{eq:Ir00s}), (\ref{eq:Ir0s0-mmm}) and (\ref{eq:Ir00s-mmm}) for sunset integrals we have used the fact that the integrals that occur have either
\begin{equation}
m=0\qquad\mbox{OR}\qquad m\neq0,\,p^2=M^2\,.
\end{equation}
The relations for the former type of integrals are valid off-shell and for the latter only on-shell.

Employing the relations given in (\ref{eq:Ir0s0}), (\ref{eq:Ir00s}), (\ref{eq:Ir0s0-mmm}), (\ref{eq:Ir00s-mmm}) and (\ref{eq:tadpole-rel}) we end up, for the case of massive external legs, with the integrals
$$I^{0000}_{mmm}(p)|_{p^2=m^2}=\frac{1}{64m^2}\qquad\textrm{and}\qquad T_m^{00}$$
and, in the case of massless external legs,
$$I^{1000}_{0mm}(p)|_{p^2=0}=\frac{p_+}{16\pi^2m^2}\,\qquad I^{0100}_{0mm}(p)|_{p^2=0}=\frac{p_-}{16\pi^2m^2}\qquad\textrm{and}\qquad T^{00}_m\,,$$
where we have evaluated the finite sunset integrals by standard means (e.g. Feynman parametrization). The tadpole integral is IR-finite but contains a logarithmic UV-divergence and it can be evaluated for example in dimensional regularization. However as the contribution from this integral cancels out in the final answer we do not need to explicitly evaluate it.

After this discussion of the regularization issues involved we can now turn to the actual computation of the two-loop correction to the dispersion relation. Since some of the expressions involved are very long we have chosen to include the full set of contributing integrals for the bosons in $AdS_5\times S^5$ and the massless bosons in $AdS_3\times S^3\times T^4$ in the appendix only. The analysis for the remaining cases is similar and we will skip most of the technical details. We start by analyzing the correction for massive modes.

\subsection{Massive modes}
The simplest case is that of $AdS_5 \times S^5$ where all worldsheet excitations have the same mass. Due to the maximal supersymmetry of the background the Green-Schwarz action is known to all orders in fermions and is given by the supercoset model of \cite{Metsaev:1998it,Arutyunov:2009ga}. Since we know in particular the sixth order $\Theta^6$-terms this allows us to compute the two-loop correction to the two-point functions for the fermions as well. Thus, for $AdS_5\times S^5$ we will compute the two-loop correction to the dispersion relation for both bosons and fermions. 

\subsubsection*{$AdS_5\times S^5$}
We start by discussing the sunset diagrams. The expression for the sunset contributions, before any integral identities are used, is quite lengthy. For completeness it is given, for the bosons, in equation (\ref{eq:full-ads5-sun}). After going on-shell and using the identities for sunset integrals derived in the previous section it simplifies dramatically however and the end result is, for bosons and fermions respectively,
\begin{equation}
\mathcal{A}_{sun}
%=-\frac{16}{3}p_1^4 I^{00,00}_{111}-\frac{1}{8}\big((2+24p_1^2)T^{00}_1T^{00}_1+6 T^{00}_1 T^{11}_1+13 T^{11}_1T^{11}_1\big) \\ \nn
%&=&-\frac{1}{12}p_1^4+\frac{3}{32}\gamma(\epsilon)^2\big(7+8p_1^2\big), \\ \nn
=
-\frac{16}{3g^2}p_1^4I^{0000}_{111}(p)
-\frac{3}{2g^2}(7+8p_1^2)[T^{00}_1]^2\,,
\quad
\mathcal{F}_{sun}
%=-\frac{16}{3}p_1^4 I^{00,00}_{111}+\frac{1}{128}\big(2(9-10p_1^2+22 \sqrt{1+p_1^2}\, p_1)T_1^{00} T_1^{00} \\ \nn
%&& -(67+28p_1^2+44 \sqrt{1+p_1^2}\, p_1)T_1^{00}T_1^{11}+T_1^{11}T_1^{11}\big)=-\frac{1}{12}p_1^4+\frac{3}{32}\gamma(\epsilon)^2 \big(1+p_1^2\big)
=
-\frac{16}{3g^2}p_1^4I^{0000}_{111}(p)
-\frac{3}{2g^2}(1+p_1^2)[T_1^{00}]^2\,.
\end{equation}
Using the integral identities for bubbles and tadpoles the bubble-tadpole contribution becomes (the full off-shell expression for the bosons without using any integral identities is given in (\ref{eq:full-ads5-bt}))
\bea
\mathcal{A}_{bt}
%-\big(T^{11}_1((1+4p_1^2)B^{00}_1-B^{11}_1)-T^{00}_1((1+4p_1^2)B^{11}_1-B^{22}_1)\big)=-\gamma(\epsilon)^2p_1^2, \\ \nn
=\frac{16p_1^2}{g^2}[T_1^{00}]^2=\mathcal{F}_{bt}\,.
%=-4p_1^2\big(T^{11}_1 B^{00}_1-T^{00}_1 B^{11}_1\big)=-\gamma(\epsilon)^2p_1^2
\eea 
so the bubble-integrals left after using (\ref{eq:triv-identity}) cancel out. 

Finally the six-vertex tadpole contribution becomes (again the full expression for the bosons is given in (\ref{eq:full-ads5-t6}))
\bea
\mathcal{A}_{t_6}
%=\frac{1}{8}\big((6-8p_1^2)T^{00}_1T^{00}_1+20T^{00}_1T^{11}_1-5T^{11}_1T^{11}_1\big)=-\frac{1}{32}\gamma_1(\epsilon)^2\big(21-8p_1^2\big)
=\frac{1}{2g^2}\big(21-8p_1^2\big)[T_1^{00}]^2\,,\qquad
\mathcal{F}_{t_6}
%=\frac{1}{8}\big((6-29p_1^2)T_1^{00}T_1^{00}-3 T_1^{00}T_1^{11}\big)=-\frac{1}{32}\gamma_1(\epsilon)^2\big(3-29p_1^2\big)
=\frac{1}{2g^2}\big(3-29p_1^2\big)[T_1^{00}]^2\,.
\eea 
Summing the three contributions gives 
\bea
\mathcal{A}_{sun}+\mathcal{A}_{bt}+\mathcal{A}_{t_6}=\mathcal{F}_{sun}+\mathcal{F}_{bt}+\mathcal{F}_{t_6}=-\frac{p_1^4}{12g^2}\,,
\eea 
in agreement with the proposed exact dispersion relation. It is worth pointing out that this is the first two-loop computation ever performed utilizing the full BMN-string so it is gratifying to see that the final result is manifestly finite and in agreement with what we expect based on symmetries and related arguments. 

\subsubsection*{$AdS_3\times S^3\times T^4$}
This theory is more complicated since beside massive worldsheet excitations, with mass $m=\hat q$ (which reduces to $1$ in the pure RR case), we now have massless excitations as well. The massless modes correspond to supersymmetries broken by the background and to describe them properly we have to use the full GS-string which currently is only known up to quartic order in fermions. This means that we can only compute the correction to the two-point function for the bosons. 

For the sunset diagrams we now have two combinations of virtual particles running in the loop; either all three particles are massive or one is massive and two are massless. The contribution of the latter turn out to exactly cancel, as can be verified from the full expression for their contribution given in (\ref{eq:ads3-massive-massless}) by making use of the identities (\ref{eq:Ir0s0}) and (\ref{eq:Ir00s}). Hence, in our regularization scheme, we see that the massless modes completely decouple from the massive dispersion relation, i.e. we could simply put them to zero in the Lagrangian and work with the supercoset sigma-model. This decoupling has already been observed for the two-loop dispersion relation in the NFS-limit and for the one-loop BMN S-matrix in \cite{Sundin:2014ema,Roiban:2014cia}.

Using the identities (\ref{eq:Ir0s0-mmm}) and (\ref{eq:Ir00s-mmm}) one finds the result
\begin{align}
\mathcal{A}_{sun}
%&\\ \nn 
%&=&
%-\frac{16}{3}\hat q^4 \big(p_1\pm q\big)^4I^{00,00}_{mmm} 
%+\frac{1}{2}\big(T^{00}_m(-2T^{11}_0+(1+12p_1^2)T_m^{00}-4T^{11}_m)-2T^{00}_0(T_0^{11}+T_m^{00}-T_m^{11})\big) \\ \nn
%&=& -\frac{1}{12}\hat q^2 \big(p_1\pm q\big)^4 +\frac{3}{8}\gamma(\epsilon)^2\big(1-4p_1^2\big)
=&
-\frac{16}{3g^2}\hat q^4\big(p_1+ q\big)^4I^{0000}_{\hat q\hat q\hat q}(p)
%+\frac{1}{2g^2}(1+12p_1^2-4\hat q^2)[T_m^{00}]^2
%-\frac{q^2}{g^2}T^{00}_\mu T_m^{00}\,. 
\\ \nn
&{}-\frac{\hat q^2}{8g^2}\big(12-161 q^2+173 q^4+72q(-2+3q^2)p_1 + 4(-12+19q^2)p_1^2\big)[T_{\hat q}^{00}]^2\,.
\end{align}
For the bubble-tadpole diagrams we find
\bea
\mathcal{A}_{bt}
%=T_m^{11}\big((1+4p_1^2)B_m^{00}-B_m^{11}\big)-T^{00}_m\big((1+4p_1^2)B_m^{11}-B_m^{22}\big)=\gamma(\epsilon)^2p_1^2
=
-\frac{4\hat q^4}{g^2} (q+p_1)^2[T^{00}_{\hat q}]^2\,.
\eea 
In the end we find that all the bubble integrals cancel among themselves and we are only left with the tadpole-type integral above. 

For the six-vertex diagram we find the following combination of tadpole integrals
\bea
\mathcal{A}_{t_6}
%= -2p_1^2 T_m^{00}T_m^{00}+T_m^{00} T_m^{11}+\frac{1}{2}T_1^{11}T_1^{11}=\frac{1}{8}\gamma(\epsilon)^2 \big(-3+4p_1^2\big)
=\frac{\hat q^2}{8g^2}\big(3(4-43q^2+47q^4)+8q(-10+19q^2)p_1+4(-4+11q^2)p_1^2\big)[T_{\hat q}^{00}]^2\,.
\eea 
Adding all the contributions together then gives
\bea \label{eq:2p-ads3}
\mathcal{A}=-\hat q^2\frac{(p_1\pm q)^4}{12g^2}\,,
\eea 
where the $\pm$ originates from whether we consider $\langle \bar z z \rangle,\langle \bar y y \rangle$ or $\langle z \bar z \rangle,\langle y \bar y \rangle$ propagators and is related to the redefinition which makes the quadratic action Lorentz-invariant, see footnote \ref{foot:q-redef}. While the final answers only differ in the sign of $q$, the intermediate steps for the tadpoles are different and hence we only presented the details for one of the cases above. 

The above is in agreement with the proposed exact dispersion relation for this case, see the introduction.

\subsubsection*{$AdS_2\times S^2\times T^6$}
For the massive sunset diagrams we have the following combinations of virtual particles in the loop: All massive, all massless or one massive and two massless. Again 
the latter two contributions cancel out so that the massless modes decouple from the calculation and one finds
\bea
& \mathcal{A}_{sun}& 
%\\ \nn
%&=&-\frac{1}{4}\frac{16}{3}p_1^4I^{00,00}_{111}+\frac{1}{4}\big(T^{00}_1(-6T^{11}_0+(1+8p_1^2)T_1^{00}-4T^{11}_1)-6T^{00}_0(T_0^{11}+T_1^{00}-T_1^{11})\big) \\ \nn
%&=&-\frac{1}{4}\frac{1}{12}p_1^4+\frac{1}{16}\gamma(\epsilon)^2\big(3-8p_1^2\big)
=-\frac{4}{3g^2}p_1^4I^{0000}_{111}(p)+\frac{1}{4g^2}(8p_1^2-3)[T^{00}_1]^2\,.
\eea 

For the bubble-tadpoles and six-vertex tadpoles we find
\bea
\mathcal{A}_{bt}
%=\frac{1}{4}\big(T_1^{11}((1+4p_1^2)B_1^{00}-B_1^{11})-T^{00}_1((1+4p_1^2)B_1^{11}-B_1^{22})\big)=\frac{1}{4}\gamma(\epsilon)^2p_1^2.
=-\frac{p_1^2}{g^2}[T^{00}_1]^2\,,
\qquad
\mathcal{A}_{t_6}
%&=&\frac{1}{8}\big(12 T_0^{11}T_0^{11}+12T_0^{11}(T_1^{00}+T_1^{11})-T_1^{00}((3+8p_1^2)T_1^{00}-9T_1^{11})\big)\\ \nn
%&=&-\frac{1}{16}\gamma(\epsilon)^2\big(3-4p_1^2\big)
=\frac{1}{4g^2}(3-4p_1^2)[T_1^{00}]^2\,,
\eea
where, again, the tadpoles with massless particles cancel among themselves. 
 
Adding the contributions together gives the manifestly finite result
\bea
\mathcal{A}=-\frac{p_1^4}{48g^2}\,,
\eea 
which differs from $AdS_5$ and $AdS_3$ only by $g\rightarrow 2g$. Similar factors of two were observed earlier in \cite{Abbott:2013kka,Roiban:2014cia,Sundin:2014ema}.

\subsection{Massless modes}
Here we present the two-loop correction to the dispersion relation for massless bosons arising from the toroidal directions of $AdS_3\times S^3\times T^4$ and $AdS_2\times S^2\times T^6$. 

\subsubsection*{$AdS_3\times S^3\times T^4$}
%Since we will find a disagreement with the exact result we will present more of the technical details in this section. 
For the sunset diagrams we have two types of contributions, either all particles running in the loops are massless or one is massless and two are massive. The full expression, i.e. off-shell and before using any integral identities can be found in eq. (\ref{eq:full-ads3-sun}). Using the identities we have derived for sunset integrals eqs. (\ref{eq:Ir0s0}) and (\ref{eq:Ir00s}) one can show that the contributions of the first type cancel out completely and we are left only with the contribution from integrals with two massive modes in the loops. Using again the identities for these integrals and taking the IR-regulator $\mu\rightarrow0$ at the end one finds the result
\bea \nn
\mathcal{A}_{sun}
%&=&-\frac12\hat q^4\big( p_-^3 I^{01,00}_{0 mm}+p_+^3 I^{10,00}_{0 mm}\big)+\big(p_+^2-p_+p_-+p_-^2 \big)T_m^{00}T_m^{00}+\mathcal{O}(\mu) \\ 
%&=&-\frac{\hat q^2}{2\pi^2}p_1^4-\frac{1}{4}\gamma(\epsilon^2)\big(p_+^2-p_+p_-+p_-^2\big)
=
-\frac{1}{2g^2}\hat q^4\big(p_+^3I^{1000}_{0\hat q\hat q}+p_-^3I^{0100}_{0\hat q\hat q}\big)
+(4-11q^2+7q^4) p_1^2 [T_{\hat q}^{00}]^2\,.
\eea 
%where we put the finite part of the amplitude on-shell in the last step and assumed that $p_1 > 0$. 

For the bubble-tadpole and six-vertex diagrams we find
\bea 
\mathcal{A}_{bt}
%=(p_+^2+p_-^2)\big(T^{11}_mB^{00}_m-T^{00}_mB^{11}_m\big)+\mathcal{O}(\mu^2)=\frac{1}{4}\gamma(\epsilon)^2 \big(p_+^2+p_-^2\big)
=
-4 \hat q^4 p_1^2 [T_{\hat q}^{00}]^2\,,
\qquad
\mathcal{A}_{t_6}= 3q^2 \hat q^2 p_1^2 [T_{\hat q}^{00}]^2\,.
\eea 
Summing all the contributions gives
\bea
\mathcal{A}_{sun}+\mathcal{A}_{bt}+\mathcal{A}_{t_6}=-\hat q^2\frac{p_1^4}{2\pi^2g^2}\,.
\eea
We see that this result differs from the proposed exact result by a factor of $6/\pi^2$, as earlier noticed in the NFS limit in \cite{Sundin:2014ema}. Furthermore note that this value is on-shell and we can not address whether the massless bosons receive a non-trivial wave-function renormalization at the two-loop level. This is because in the derivation we encounter sunset integrals with $r,u\neq 0$ and $s,t\neq 0$ and in order to disentangle the powers of loop momenta we need to use (\ref{eq:Ir00s}). However, that relation is derived using the on-shell assumption $p^2 = m^2$ so naturally the final answer is also on-shell.

As a final comment we point out that the discrepancy with the exact proposal can not be explained by a potential inadequacy of the regularization procedure. This can be understood from the fact that the tensor-structure of the integrals forces them to be proportional to appropriate powers of masses and external momenta $p_\pm$. Using this in (\ref{eq:full-ads3-sun}) and going on-shell shows, without using any integral identities at all, that the only integrals contributing to the final piece of the amplitude indeed are the ones listed above.

\subsubsection*{$AdS_2\times S^2\times T^6$}
For the sunset diagrams in the massless sector of $AdS_2\times S^2\times T^6$ we have either two massive and one massless, one massive and two massless or three massless virtual particles propagating in the loops. Again the latter two contributions cancel out and the result comes only from the diagrams with two massive modes in the loops. One finds
\bea
\mathcal{A}_{sun}
%=-\frac{1}{4}p_-^3I_{0mm}^{01,00}+\frac{1}{4}(p_+-p_-)^2T_1^{11}T_1^{00}=-\frac{1}{4\pi^2}p_1^4-\frac{1}{16}\gamma(\epsilon)^2\big(p_+-p_-\big)^2
=-\frac{1}{4g^2}\big(p_-^3I_{011}^{0100}+p_+^3I_{011}^{1000}\big)+\frac{1}{g^2}p_1^2[T_1^{00}]^2\,.
\eea 
while for the bubble-tadpole diagram we have
\bea
\mathcal{A}_{bt}
%=\frac{1}{4}(p_+^2+p_-^2)\big(T_1^{11}B_1^{00}-T_1^{00}B_1^{11}\big)+\mathcal{O}(\mu^2)=\frac{1}{16}\gamma(\epsilon)^2\big(p_+^2+p_-^2\big)
=-\frac{1}{g^2}p_1^2[T_1^{00}]^2\,,
\eea 
and the six-vertex tadpoles give a vanishing contribution.
%\bea
%\mathcal{A}_{t_6}
%=\frac{1}{2}p_+p_-\big(T_1^{00}-T_0^{11}\big)+\mathcal{O}(\mu^2)=-\frac{1}{8}\gamma(\epsilon)^2 p_+p_-
%=0
%\eea 
Summing the various contributions again yields a manifestly finite result, 
\bea
\mathcal{A}_{sun}+\mathcal{A}_{bt}+\mathcal{A}_{t_6}=-\frac{p_1^4}{4g^2\pi^2}
\eea
where we note that the relative difference between the massless propagators of $AdS_2$ and $AdS_3$ is a factor of two, as compared with the massive sector where the relative difference was a factor of four.

\section{Massless S-matrix in $AdS_3\times S^3\times T^4$}
\label{sec:massless-S}
Having performed a detailed analysis of two-point functions and dispersion relations we now turn to a study of the massless S-matrix for two-to-two scattering in order to try to clarify the role of the massless modes. As we will see up to one loop our results are consistent with the symmetries and the one-loop dressing phase in the massless sector is simply given by the massless limit of the BES/HL phase (up to an IR-divergent piece arising as an artifact of the massless limit). 

In fact we are able to go further and also compute an element of the two-loop S-matrix. The S-matrix element we consider is a forward type scattering element of a pair of $u_1$ massless bosons and we will show that once we go on-shell the entire amplitude vanishes. Naively this is a very involved computation since many of the contributing integrals are very hard to evaluate. Luckily we never need to tackle this problem since it is enough to know the IR-scaling, together with a few algebraic identities, to argue that on-shell the entire contribution vanishes.

\subsection{Massless S-matrix to one loop}
To simplify the analysis we will work with the $AdS_3\times S^3\times T^4$ background supported by pure RR-flux, i.e. $q=0$. In the following the momenta of the incoming particles will be denoted $p$ and $q$, where the latter should not be confused with the parameter $q$ of the background. Schematically a two-particle massless scattering amplitude takes the form
\bea
\mathbbm{T} \ket{A(p) B(q)} = \sum_{C,D} T^{CD}_{AB} \ket{C(p) D(q)}\,,\qquad p_1>0>q_1\,,
\eea 
where we used the standard notation $\mathbbm{S}=\mathbbm{1}+i \mathbbm{T}$ and have assumed that the transverse momenta $p_1$ and $q_1$ have opposite sign.

\begin{table}[ht]
\begin{center}
\begin{tabular}{ccccccccc}
& $y$ & $z$ & $u_1$ & $u_2$ & $\chi^1$ & $\chi^2$ & $\chi^3$ & $\chi^4$ \\
\hline
\vphantom{${}^{A^{A^a}}$}
$U(1)_1$ & $-1$ & $0$ & $0$ & $0$ & $-1/2$ & $1/2$ & $1/2$ & $-1/2$  \\
$U(1)_2$ & $0$ & $-1$ & $0$ & $0$ & $1/2$ & $-1/2$ & $1/2$ & $-1/2$ \\
$U(1)_3$ & $0$ & $0$ & $-1$ & $0$ & $1/2$ & $1/2$ & $-1/2$ & $-1/2$
\end{tabular}
\end{center}
\caption{Summary of $U(1)$ charges for $AdS_3\times S^3\times T^4$.}
\label{tab:charges-ads3}
\end{table}

Denoting the four massless bosonic excitations $u_i$ and $\bar u_i$, with $i=1,2$, we find that the only non-zero processes at tree-level, with purely bosonic in-states, are\footnote{There are no massless to massive or massive to massless S-matrix elements, as we have verified explicitly at tree-level. This in fact follows from the classical integrability of the full string (including massless fermions) established in \cite{Sundin:2012gc,Wulff:2014kja}.}
\bea 
&& \mathbbm{T}^{(0)}\ket{u_2(p) u_2(q)}=\frac{i}{g} \ell_0 \big(\ket{\bar \chi_3(p) \chi_3(q)}+\ket{ \bar \chi_4(p) \chi_4(q)}\big)\,, \\ \nn
&& 
\mathbbm{T}^{(0)}\ket{u_1(p) u_2(q)}=\frac{1}{g} \ell_0 \big(\ket{\chi_3(p)   \chi_4(q)}-\ket{\chi_4(p) \chi_3(q)}\big)\,, \\ \nn
&&
\mathbbm{T}^{(0)}\ket{u_1(p) \bar u_1(q)}=-\frac{i}{g} \ell_0 \big(\ket{\chi_3(p) \bar \chi_3(q)}+\ket{  \chi_4(p) \bar\chi_4(q)}\big)\,, \\ \nn
&&  \mathbbm{T}^{(0)} \ket{u_2 (p)\bar u_1(q)}=-\frac{1}{g} \ell_0 \big(\ket{\bar \chi_3(p)  \bar  \chi_4(q)}-\ket{\bar \chi_4(p) \bar \chi_3(q)}\big)\,,
\eea 
where we stress that we are looking at the kinematic region where $p_1>0>q_1$ and we have not written out elements related by complex conjugation. We see that there is no manifest symmetry in exchanging $u_1$ and $u_2$ which is a consequence of working with the type IIA version of the background. In fact, the last massless boson $u_2$ is neutral under the global $U(1)$'s, see table \ref{tab:charges-ads3}, and this allows distinct processes as compared to those with $u_1$ in-states.

The contributing diagrams at one loop are a sum of four-vertex $s,t$ and $u$-channel topologies
\bea \label{diagram:stu}
\parbox[top][0.8in][c]{1.2in}{\fmfreuse{schannel}}+\parbox[top][0.8in][c]{1.2in}{\fmfreuse{tchannel}}
+\parbox[top][0.8in][c]{1.2in}{\fmfreuse{uchannel}}
%+\parbox[top][0.8in][c]{1.5in}{\fmfreuse{tadpolesix}}
\eea
and a six-vertex tadpole
\bea
\label{diagram:t6}
\parbox[top][0.8in][c]{1.5in}{\fmfreuse{tadpolesix}}
\eea 
Computing the diagrams we find that the final answer is manifestly UV-finite and that only diagonal scattering processes are non-zero. Generically we have\footnote{We have not computed the (divergent) tadpole contribution for $FF\rightarrow FF$ amplitudes since these would need the $\Theta^6$-terms in the Lagrangian which are not currently known.}
\bea
\mathbbm{T}^{(1)}\ket{A_i(p) B_j(q)}=\frac{1}{g^2} \ell_1 \ket{A_i(p)B_j(q)}+\textrm{possible real terms}\,,
\eea
where $A$ and $B$ can be any state, bosonic or fermionic. Up to the real terms, which are fully determined from tree-level elements via the optical theorem, we have computed all possible scattering elements for one-loop amplitudes. Note that in order to evaluate the one-loop amplitudes we have repeatedly made use of the identities in (\ref{eq:triv-identity}) and (\ref{eq:B00-rel}) which allow us to rewrite any divergent bubble-type integral in terms of tadpoles and $B^{00}$.

The explicit form of the tree-level and one-loop amplitudes are
\bea
\label{eq:massless-amp}
\ell_0 =\frac{1}{2} \sqrt{p |q|}\,,\qquad 
\ell_1=\frac{i}{4\pi}\big(1-\log 4 p |q|\big)p |q|\,.
\eea
Here we have included the overall Jacobian from the energy-momentum delta-function and the external leg factors given by
\bea \nn
J_E = -\frac{1}{8 p q}+\mathcal{O}(g^{-2})\,.
\eea 
As observed in \cite{Sundin:2013ypa}, the one-loop amplitude is manifestly finite by consequence of non-trivial cancellations between four- and six-vertex Feynman diagram topologies.

\subsection{S-matrix element for massless modes at two loops}
\label{section:2loopS}
We now go to the two-loop level, again in $AdS_3\times S^3\times T^4$ with zero NSNS-flux, and compute the amplitude
\bea\label{eq:massless-amp-2loop}
\mathcal{A}\big(u_1(p) u_1(q)\rightarrow u_1(p)u_1(q)\big),\qquad (p_1 >0>q_1)\,.
\eea 
We choose to look at this specific element since the relevant vertices in the Lagrangian are relatively simple. Furthermore, since the tree-level amplitude for this process is zero, we can neglect any two-loop wave function renormalization, which we have not determined. Nevertheless, the number of terms that appear in the actual computation is very large and we will be rather brief in technical detail in this section. 

At two loops we need to expand the Lagrangian to $\mathcal{O}(g^{-3})$, i.e. to eight order in transverse fluctuations and thus a large class of distinct Feynman diagrams contribute. Furhermore, the eight order Lagrangian is rather complicated to derive but luckily the relevant terms that can contribute to the amplitude we are looking at are simple. The contributing quartic $u_1$-terms that appear in the bosonic Lagrangian are given by
\bea
\label{eq:quarticy3}
\mathcal{L}=-\frac{1}{4 g^2}|\partial_+u_1\partial_-u_1|^2\big(|z|^2+|y|^2\big)+\frac{1}{2 g^3}(|\partial_+ u_1|^4-|\partial_-u_1|^4)\big(|z|^4-|y|^4\big)+\dots
\eea 
These are in fact all the terms quartic in $u_1$. Naively one would expect to also get additional quartic $u_1$-terms involving fermions from $\mathcal{L}^{(2)}$. To see that these are in fact absent note that prior to the gauge-fixing, the $BF$ part of the Lagrangian has terms quadratic in $u_1$ of the form
\bea
\label{eq:Lbf4y3}
\mathcal{L} =
\frac12\gamma^{+-}\partial_-u_1\partial_+\bar u_1\big(\chi^-_4 \bar\chi_4^+ +\chi^+_3 \bar \chi^-_3\big) +\mbox{h.c}+\mathcal{O}(\chi^4)\dots
\eea 
Fixing the gauge (\ref{eq:gf1}) introduces higher order corrections to the worldsheet metric as
\bea
\gamma^{ij} = \eta^{ij} + \frac{1}{g} \gamma_2^{ij} + \frac{1}{g^2}\gamma_4^{ij}+\frac{1}{g^3}\gamma_6^{ij}+\dots
\eea 
where the subscript denotes the number of transverse fluctuations in each term. The leading order part, $\gamma_2^{ij}$, only depends on the massive coordinates while the higher orders have, at most, terms quadratic in $u_1$.\footnote{This explains why we did not have to write out the $\mathcal{O}(\chi^4)$-terms in (\ref{eq:Lbf4y3}), while these can be quadratic in the massless bosons they can not induce any massless (bosonic) vertices at $g^{-3}$.} The condition $\det\gamma^{ij}=-1$ implies
\bea \nn
\gamma^{11}=-1+\frac{1}{g}\gamma_2^{00}+\frac{1}{g^2}\big(-(\gamma^{00}_2)^2 + \gamma_4^{00}\big)+\dots
\eea 
but since $\gamma^{+-}=\frac14\big(\gamma^{00}-\gamma^{11}\big)=\frac12+\frac{1}{4 g^2}\big(\gamma_2^{00}\big)^2+\mathcal{O}(g^{-3})$ we see that the dependence on the higher order correction $\gamma_4^{00}$ drops out and therefore we only have purely bosonic quartic $u_1$-terms in $\mathcal{L}_8$.

\subsubsection*{Four-vertex diagrams}
We start the discussion with the most complicated diagrams which are of the wineglass type.
\begin{figure}[!htb]
\centering
\includegraphics[scale=.7]{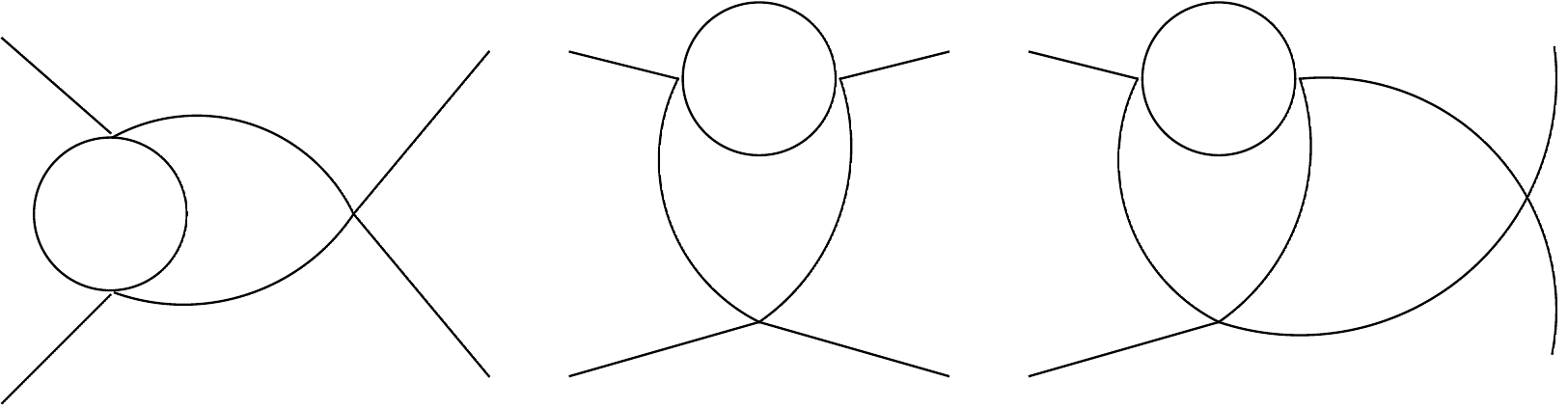}
\caption{Three out of the six wineglass diagrams. The other three are obtained by flipping the position of the bubble loop.}
\label{fig:wine}
\end{figure} 
For the amplitude (\ref{eq:massless-amp-2loop}) only $t$ and $u$-channel diagrams are non-zero and the explicit structure of the quartic string Lagrangian is such that the virtual particles appearing in the loops have mass combinations: $\{0000\},\{1100\},\{0111\}$. Thus, the integrals that appear are of the form
\bea
\nn
W^{rstu}_{m_1m_2m_3}(p,q) =\int\frac{d^2kd^2 l}{(2\pi)^4}\frac{k_+^r k_-^s l_+^t l_-^u}{(k^2-m_1^2)(l^2-m_2^2)((k+l-p)^2-m_3^2)((k+l-q)^2-m_3^2)}\,.
\eea
%Before imposing overall energy-momentum conservation 
The $t$-channel integrals depend on two external momenta $p,p'$ or $q,q'$ scaling like
\bea \label{eq:on-shell-t}
p_-\,,p'_- \sim \mu^2\,,\qquad\textrm{or}\qquad  q_+\,,q'_+ \sim \mu^2\,,
\eea
in the kinematic regime $p_1 > 0 > q_1$ where $\mu$ is the IR-regulator mass. Using this simple fact one can deduce that integrals with massive virtual particles have at most a $1/\mu^2$ IR-divergence in the small mass-limit while integrals with only massless virtual particles scale as $1/\mu^4$ (see \cite{Klose:2007rz} for explicit expression for some of the wineglass integrals). 

Putting the external momenta on-shell one finds that the only integrals that can contribute are those with massless virtual particles. Using the symmetry of the remaining integrals we can furthermore write
\bea
\label{eq:W-scaling}
W^{rstu}_{\mu\mu\mu}(p,p') = \alpha_{rstu}(p_++p'_+)^{r+t}(p_-+p_-')^{s+u}W^{0000}_{\mu\mu\mu}(p,p')\,,
\eea
where $\alpha_{rstu}$ are some, possibly $1/\epsilon$-dependent, constants. Using this to rewrite any integral with non-zero powers $r,s,t$ and $u$ together with (\ref{eq:on-shell-t}) we find that the entire expression vanishes once we send $\mu\rightarrow 0$. That is, on-shell the entire $t$-channel contribution is zero!

For the $u$-channel the IR-divergence of each class of integrals is less severe. The maximal divergence is $1/\mu^2$ and again the only non-trivial contribution, i.e. which does not immediately disappear in the $\mu\rightarrow 0$ limit, are integrals with only massless propagators, i.e. $W^{rstu}_{\mu\mu\mu}(p,q)$. However, since the integral depends on both $p$ and $q$ and thus mixes right- and left-moving momenta, using identities similar to (\ref{eq:W-scaling}) does not immediately put the $u$-channel contribution to zero, naively it looks like there remains a $1/\mu$ divergence. Thus the analysis becomes a bit more involved and we will need to make use of some algebraic identities for the integrals. The first one follows from (\ref{eq:triv-identity}) and reads
\bea 
\label{eq:sunset-id1}
W^{r+1,s+1,t,u}_{\mu\mu\mu}=\mu^2 W^{rstu}_{\mu\mu\mu}
+\int\frac{d^2kd^2l}{(2\pi)^4}\frac{k_+^r k_-^s l_+^t l_-^u}{(l^2-\mu^2)((k+l-p)^2-\mu^2)((k+l-q)^2-\mu^2)}\,.
\eea
The last integral reduces to a sum of bubble times tadpole integrals by shifting the loop momentum variable $k\rightarrow p-k-l$. This identity can be used to reduce all wineglass-integrals to integrals with powers of only $k_\pm^r l_\pm^s$ or $k_\pm^r l_\mp^s$ plus bubble and tadpole integrals.

The wineglass integrals that mix left- and right-moving loop momenta, $k_\pm^r l_\mp^s$, can be further reduced using (we're suppressing the argument of $W$)
\begin{align}
W^{r00s}_{\mu\mu\mu}=&I^{r-1,0,0,s-1}_{\mu\mu\mu}(q)
-(p^2-\mu^2)W^{r-1,0,0,s-1}_{\mu\mu\mu}
-W^{r-1,1,1,s-1}_{\mu\mu\mu}
-W^{r,1,0,s-1}_{\mu\mu\mu}-W_{\mu\mu\mu}^{r-1,0,1,s}
\nn\\
&{}
+p_+\big(W^{r-1,1,0,s-1}_{\mu\mu\mu}+W^{r-1,0,0,s}_{\mu\mu\mu}\big) 
+p_-\big(W^{r,0,0,s-1}_{\mu\mu\mu}+W^{r-1,0,1,s-1}_{\mu\mu\mu}\big)
%, \\ \nn
%W^{0s,u0}_m&=&I^{0s-1,u-10}_m(q)+m^2 W^{0s-1,u-10}-W^{1s,u-10}_m-W_m^{0s-1,u1}+p_-\big(W^{1s-1,u-10}_m+W^{0s-1,u0} \big) \\ 
%&& +p_+\big(W^{0s,u-10}_m+W^{0s-1,u-11}_m\big)-p^2W^{0s-1,u-10}_m-W^{1s-1,u-11}
\label{eq:sunset-id3}
\end{align}
and a similar identity for $W^{0rs0}$ where $I^{rstu}_{\mu\mu\mu}(q)$ is the standard sunset-type integral in (\ref{eq:sunset-integral}). The above identity follows from rewriting numerator terms as
$$
k_+ l_- = (k+l-p)^2-m^2 + m^2 +\dots
$$
This allows us to reduce most wineglass-integrals with powers of $k_\pm l_\mp$ to powers of $k_\pm l_\pm$ and additional sunset- and tadpole-type integrals. However, at the end we are still left with $W^{0110}_{\mu\mu\mu}$ and $W^{1001}_{\mu\mu\mu}$ integrals. Luckily we can reduce these using the symmetry in $k$ and $l$ which gives
\begin{equation} 
W^{0110}_{\mu\mu\mu}=
% \\ \nn && 
W^{1001}_{\mu\mu\mu}=\frac12I^{0000}_{\mu\mu\mu}(q)-\frac{p^2-\mu^2}{2}W^{0000}_{\mu\mu\mu}-W^{0011}_{\mu\mu\mu}+p_+ W^{0001}_{\mu\mu\mu}+p_-W^{0010}_{\mu\mu\mu}\,.
\label{eq:sunset-id4}
\end{equation} 
Thus we have reduced all integrals mixing powers of $k_\pm l_\mp$ to wineglasses of powers $k_\pm l_\pm$ plus additional sunset and bubble/tadpole-type integrals. 

The reason we went through this rather lengthy reduction scheme is because now the remaining wineglass integrals drop out in the $\mu\rightarrow 0$ limit. That is, using the fact that $W \sim 1/\mu^2$ together with 
\bea
W^{rstu}_{\mu\mu\mu}(p,q)=\alpha_{rstu}( p_++q_+)^{r+t}(p_-+q_-)^{s+u}W^{0000}_{\mu\mu\mu}(p,q)
\eea
we find that each term $W^{r0s0}_{\mu\mu\mu}(p,q)$ or $W^{0r0s}_{\mu\mu\mu}(p,q)$ is multiplied by high enough powers of $p_-$ and $q_+$ to completely vanish when we take the regulator to zero. 

Since we have used the identities (\ref{eq:sunset-id1}), (\ref{eq:sunset-id3}) and (\ref{eq:sunset-id4}) we still have to see what happens with the bubble/tadpole integral in (\ref{eq:sunset-id1}) and the additional sunset-integrals. The sunset integrals are at most $1/\mu^2$ and as it turns out, they are again multiplied with high enough powers of $p_-$ and $q_+$ to completely vanish once we take $\mu$ to zero. The only terms left are thus the bubble/tadpole-type integrals in (\ref{eq:sunset-id1}). Naively one would think these scale like $1/\mu^2$ but this is not true in the $u$-channel. The reason is that bubble integrals only have a $\log\mu$ divergence in the regime where $p_1>0>q_1$,
\bea \nn
B^{rs}_\mu(p_+,q_-) \sim \log \mu+\dots
\eea
Thus we have that, at most, the second term on the RHS of (\ref{eq:sunset-id1}) goes as $(\log \mu)^2$. Looking at the specific terms appearing we again see that they are multiplied with high enough powers of the external momenta to vanish once we take the regulator to zero. We can therefore conclude that
\bea
\mathcal{A}_{wineglass} = 0
\eea
without having to explicitly evaluate any wineglass type integrals (which is very complicated). 

The next class of diagrams to consider are the so-called double bubble diagrams, involving two standard bubble-type integrals. 
\begin{figure}[!htb]
\centering
\includegraphics[scale=.7]{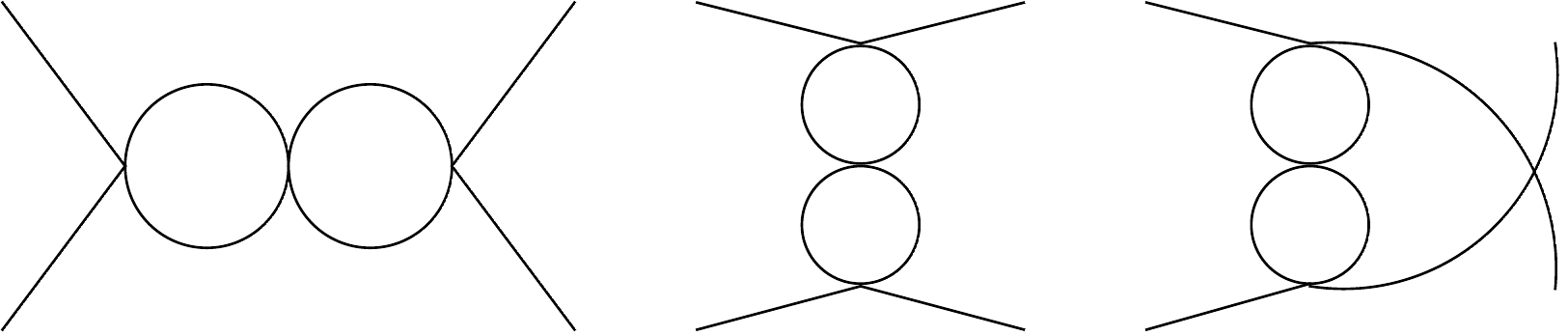}
\caption{Double bubble diagrams}
\label{fig:bubbub1}
\end{figure} 
Since we do not have any quartic vertices consisting of massless bosons alone, the $s$-channel contribution to the (\ref{eq:massless-amp-2loop}) process is again trivially zero. For both $t$ and $u$-channels we find that the bubble integrals that appear enter as
\bea \nn
B_1^{rs}(P) B_0^{tu}(P'), \qquad B_0^{rs}(P) B_0^{tu}(P')
\eea
where $P$ and $P'$ are some (possibly zero) combination of external momenta. The massive bubble integral is at most $\log \mu$ divergent, while the massless one is $1/\mu^2$ in the $t$-channel and again $\log \mu$ in the $u$-channel. Furthermore, imposing that the external momenta are on-shell, again for $p_1>0>q_1$, we find without using any reduction schemes that the entire double bubble contribution goes to zero. Thus we conclude that also
\bea 
\mathcal{A}_{double-bubble}=0\,.
\eea

We have one additional class of diagrams built out of three four-vertices and that is the one-loop bubble diagram, depicted in (\ref{diagram:stu}), with a tadpole added on one of the internal propagators.
\begin{figure}[!htb]
\centering
\includegraphics[scale=.6]{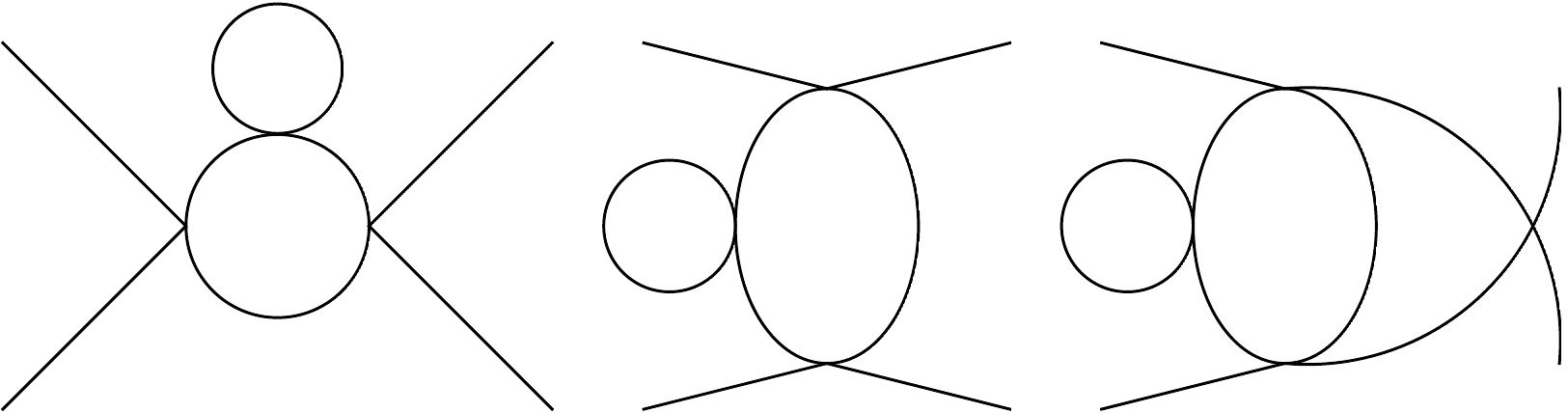}
\caption{Another double-bubble type diagram. Some trivial permutations are not written out.}
\label{fig:bubbub2}
\end{figure} 
This diagram is the two-loop generalization of the $s,t$ and $u$-channel diagrams in (\ref{diagram:stu}). The only tadpole propagators that are non-zero off-shell are the ones for the two massive bosons $z$ and $y$. However, since they come with opposite sign \cite{Roiban:2014cia}, any tadpole on an internal line will sum up to zero and we have
\bea
\mathcal{A}_{bub-tad} = 0\,. 
\eea 
It is only the diagrams built out of four-vertices alone that can give a finite contribution to the amplitude. Since we have seen that they vanish the only thing that remains to verify is that the higher vertex diagrams, which can only give divergent contributions, are also zero as we expect.

\subsubsection*{Four- and six-vertex diagrams}
\begin{figure}[!htb]
\centering
\includegraphics[scale=.6]{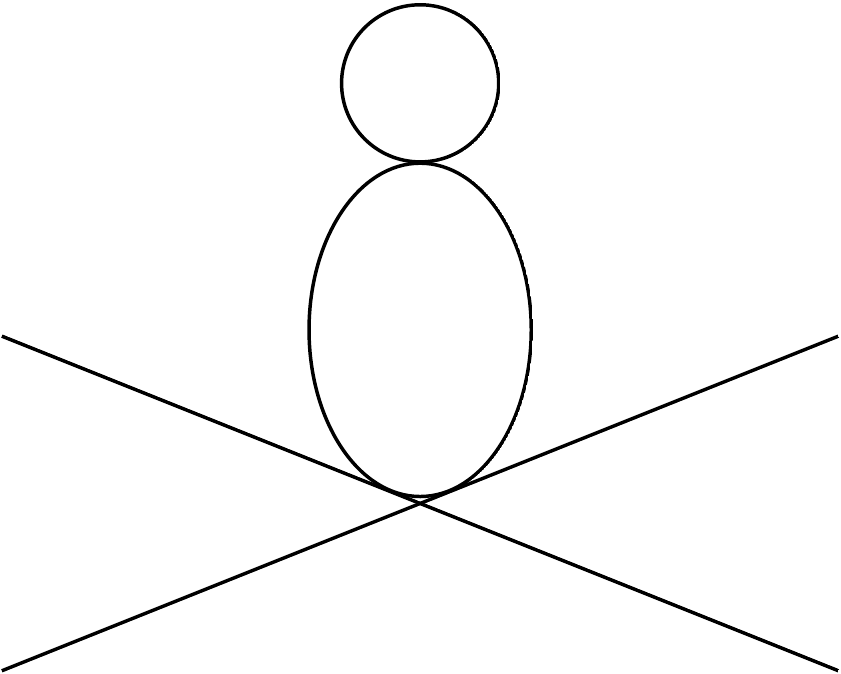}
\caption{Bubble and a tadpole with four- and six-vertices.}
\label{fig:bubtad1}
\end{figure}

First we consider the four-point generalization of (\ref{diagram:8}) which is a combination of a bubble- and tadpole-type integral, see figure \ref{fig:bubtad1}.
Since the tadpole sits on a four-vertex we can again conclude from the form of the off-shell propagator in \cite{Roiban:2014cia} that the entire contribution vanishes. 
\begin{figure}[!htb]
\centering
\includegraphics[scale=.6]{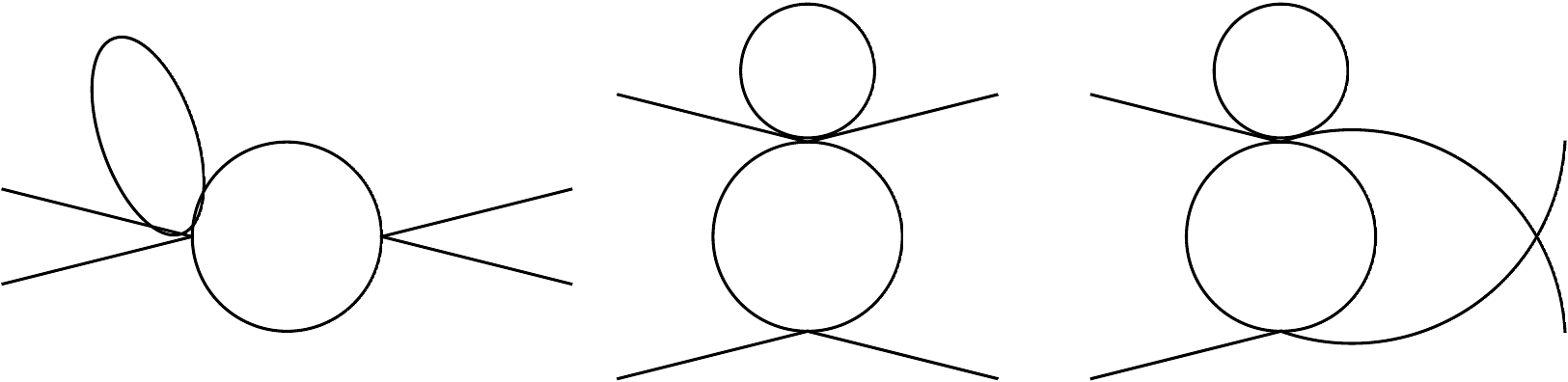}
\caption{Another bubble-tadpole type diagram. We have suppressed the three additional diagrams with the tadpole attached to the second vertex.}
\label{fig:bubtad2}
\end{figure} 

Another diagram that combines four- and six-vertices is obtained by adding a tadpole on the vertices of the $s,t$ and $u$-channel diagrams in (\ref{diagram:stu}), as depicted in figure \ref{fig:bubtad2}. As before the $s$-channel is trivially zero. Since the tadpole sits at the six-vertex we can not immediately conclude that it will sum up to zero. However, computing the explicit contribution and using the identities (\ref{eq:bubble-id1}) we can reduce the bubble integrals to $B_0^{00}$ and $B_1^{00}$. Taking the external momenta $p$ and $q$ to be on-shell we find that each class of diagram is identically zero.

With this we conclude that all diagrams mixing four- and six-vertices are zero,
\bea
\mathcal{A}_{4-6\, vertex} = 0\,.
\eea

\subsubsection*{Eight-vertex diagram}
The last diagram to consider is the double tadpole diagram constructed from a single eight-vertex. However, from (\ref{eq:quarticy3}) we see that since there is a relative sign between the massive coordinates, the contribution from tadpoles with the massive $z$ and $y$ modes will cancel out. Thus again we conclude that this diagram does not contribute to the amplitude,
\begin{figure}[!htb]
\centering
\includegraphics[scale=.4]{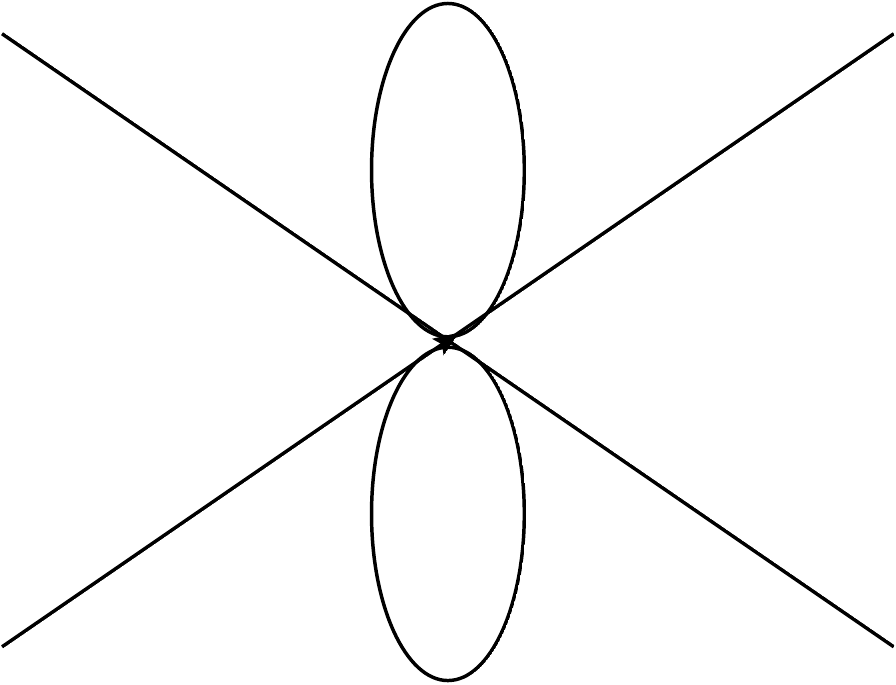}
\caption{The eight-vertex double tadpole.}
\label{fig:tad8}
\end{figure}
\bea
\mathcal{A}_{8-tad}=0\,.
\eea

\subsubsection*{Summary}
Since this section was rather brief in technical detail and involved a lot of distinct Feynman diagrams, we briefly summarize what we have found here. At the classical level we found that for incoming massless bosons only S-matrix elements with out-going fermions are non-zero. This is in qualitative agreement with \cite{Borsato:2014hja}, although a direct comparison is complicated by the fact that that work refers to the type IIB case. Furthermore, at the one-loop level we found that only diagonal processes were non-zero which indicates that we should have a non-trivial phase-factor at this order. 

At the two-loop order, where we for simplicity restricted to a single forward-type scattering element, we found that the entire contribution vanished on-shell. That is,
\bea 
\mathcal{A}\big(u_1(p) u_1(q)\rightarrow u_1(p)u_1(q)\big) = 0\,,
\eea 
in the kinematic region where $p_1>0>q_1$, or equivalently $p_-=q_+=0$. Unless there is a remarkable cancellation between scattering factors and the phase, this suggests that the phase is zero at the two-loop order.

\subsection{Comparing to S-matrix derived from symmetries: extracting the phase}
\label{sec:exact}
For the type IIB-string in $AdS_3\times S^3 \times T^4$ the form of the massless S-matrix was fixed, up to some phases, from the symmetries in \cite{Borsato:2014hja}. However, since we are working in type IIA where the RR-flux breaks the local SO(4)-invariance of the $T^4$ present in the type IIB case, it is not immediately straight-forward to compare their S-matrix to ours. Here we will simply assume that since the coordinate $u_1$ is unaffected by the T-duality involved the S-matrix element for
\bea
\nn
u_1 u_1 \rightarrow u_1 u_1
\eea
should be the same in type IIA and type IIB at least up to two loops. From the worldsheet analysis in the previous section we know that this element should have a first non-trivial term at the one-loop order, i.e., the tree-level part should be zero. 

Symmetry dictates that this amplitude be given simply by a phase \cite{Borsato:2014hja} (their eq. (5.33))\footnote{We are grateful to R. Borsato and O. Ohlsson Sax for pointing out a mistake in the form of the exact amplitude used in an earlier version of this paper.}
\bea \label{eq:sax1}
\mathcal{A}_{11} =(\sigma^{\circ\circ}_{pq})^{-2}\,.
\eea 
This phase is as of yet undetermined and we want to find it by comparing to our perturbative calculations. A natural guess might be that it is simply given by the massless limit of the standard AFS-phase \cite{Arutyunov:2004vx}. However this would imply a non-trivial S-matrix element at tree-level contradicting the perturbative results. We conclude instead that the classical part of the phase vanishes. With vanishing tree-level phase we find only a one-loop contribution to the S-matrix element
\bea
\label{eq:a11-tree}
\mathcal{A}_{11} = 1+\frac{i}{h^2}\theta_1(p,q)+\mathcal{O}(h^{-3})\,,
\eea
where $\theta_1(p,q)$ is the one-loop dressing phase.

For the massive sector of $AdS_3\times S^3\times T^4$ the one-loop phases were determined in \cite{Borsato:2013hoa,Beccaria:2012kb,Abbott:2013mpa}. There are two distinct phases, labeled with respect to the underlying symmetry groups as $LL$ and $LR$ (or equivalently $RR$ and $RL$). Writing the phases as $\theta_{LL}$ and $\theta_{LR}$ for excitations with mass $m$, the explicit expressions are given by
\bea
&& \theta_{LL}(p,q)=-\frac{1}{4\pi}\frac{p^2 q^2\big( \textbf{p}\cdot\textbf{q}+m^2\big)}{(\varepsilon_q p - \varepsilon_p q)^2}\log \frac{q_-}{p_-}+\frac{1}{8\pi}\frac{p q(p+q)^2}{\varepsilon_q p - \varepsilon_p q} \\ \nn
&& \theta_{LR}(p,q)=-\frac{1}{4\pi}\frac{p^2 q^2\big( \textbf{p}\cdot\textbf{q}-m^2\big)}{(\varepsilon_q p - \varepsilon_p q)^2}\log \frac{q_-}{p_-}-\frac{1}{8\pi}\frac{p q(p-q)^2}{\varepsilon_q p - \varepsilon_p q}
\eea 
which sum to the HL-phase, $\theta_{LL}+\theta_{LR}=\theta_{HL}$. Assuming $p_1> 0 > q_1$ and expanding around the zero mass case gives
\bea 
\theta_{LL}(p,q)=\frac{1}{8\pi} p|q| \log \frac{4 p |q|}{m^2}-\frac{1}{16\pi}\big(p+q)^2\,, \quad
\theta_{LR}(p,q)=\frac{1}{8\pi} p|q| \log \frac{4 p |q|}{m^2}+\frac{1}{16\pi}\big(p-q)^2\,,
\eea 
neither of which agrees with (\ref{eq:massless-amp}). However, taking the sum of the two phases gives
\bea 
\label{eq:sum-LL-LR}
\theta_{HL}(p,q)=\theta_{LL}(p,q)+\theta_{LR}(p,q)=-\frac{1}{4\pi} p|q| \log \frac{4 p |q|}{m^2}+\frac{1}{4\pi} p|q|\,.
\eea 
Thus we note that, up to a divergent $\log m$ term, the massless limit of the HL-phase agrees with explicit worldsheet calculations, eq. (\ref{eq:massless-amp}), i.e. $\theta_1=\theta_{HL}$. The divergent term is simply an artifact of the way we take the massless limit.

What about two loops? Since we found that the contribution to $\mathcal{A}_{11}$ at this order was zero there should not be any two-loop contribution to the phase. However it is not really justified to try to compare our results to those of \cite{Borsato:2014hja} at this order. The reason is that the dispersion relation we find for the massless modes disagree and this difference could affect the suggested exact result at this order. Ignoring this issue for the moment let us examine the massless limit of the two-loop part of the BES phase. Using \cite{Beisert:2006ez}
\bea \nn
c^{(2)}_{r,s} = \frac{1}{48}\big( 1 + (-1)^{r+s}\big)(1-r)(1-s)
\eea 
we can use this in the sum over $r,s$ to get the two-loop phase in a closed form. Plugging this in, and using $m/h$ as a prefactor in the exponent, gives
\bea 
e^{i\frac{m}{h} \theta_{2} } =1 -\frac{i}{192 h^3}\frac{p q \big(p^2+ q^2\big)}{m}
\eea
which is IR-divergent. The fact that there is no finite piece seems to be consistent with what we find but it is not obvious how to remove, in a natural way, the IR-divergent piece.

In summary it therefore appears that our findings are consistent with a massless sector dressing phase which consists of simply the massless limit of the HL-phase
\begin{equation}
(\sigma^{\circ\circ})^{-2}=e^{\frac{i}{h^2}\theta_{HL}}\,,
\end{equation}
at least up to one loop.

\section{Conclusion}
We have performed the first full two-loop calculation for $AdS_n\times S^n\times T^{10-2n}$ strings in the BMN regime. The computations build on earlier methods developed in \cite{Roiban:2014cia} where it was understood how to properly regularize the theory. First we computed the two-loop correction to the dispersion relation for massive and massless modes and compared with the exact dispersion relation of \cite{Hoare:2013lja,Lloyd:2014bsa}. While the massive sector agrees with what is expected from symmetry arguments we find a curious discrepancy with the proposed dispersion relation in the massless sector of $AdS_3\times S^3\times T^4$. The discrepancy shows that our understanding of the massless modes of the BMN sector for $n=2,3$ is still incomplete. 

To improve our understanding of the massless modes we have also calculated the S-matrix in the massless sector up to one loop. It takes a form similar to that suggested in \cite{Borsato:2014hja} based on symmetries, with a phase that is simply the massless limit of the HL-phase (with an IR-divergence subtracted to get a finite result). We also computed a single massless forward type scattering element at the two-loop level. The amplitude turned out to completely vanish once we put the external momenta on-shell, indicating that there should be no phase factor at this order in perturbation theory. 
 
There are several natural extensions of this work. Most pressing is to understand the discrepancy we find for the massless dispersion relation. As mentioned in the introduction one possible resolution is that the central charges receive non-trivial quantum corrections. One way to address this question would be to compute the (quantum) algebra of two supercharges. Technically this is a challenging problem and we might return to it in the future. 

It would also be very interesting to try to understand the massless phase(s) in more detail. The two-loop computation we performed is a first step in this direction. A natural extension would be to compute the full S-matrix at two loops. In order to do this we might be forced to evaluate the wineglass integrals properly. For the case of virtual particles with different masses this is a very challenging problem. Again, we might return to this computation in the future. 

As we mentioned in the introduction $AdS_n \times S^n \times T^{10-2n}$ only exhibit even numbered vertices in the BMN expansion. It would be interesting to extend the analysis performed in this paper to the more challenging backgrounds of, say, $AdS_4\times \mathbbm{CP}^3$ and $AdS_3\times S^3 \times S^3 \times S^1$. Several novel complications arise in these backgrounds. For example, the interpolating function $h(g)$ receives a non-trivial correction already at one loop. Furthermore, the mass spectrum is richer and there exist certain heavy (worldsheet) modes that seem to be composites of lighter excitations. While it's probably out of the question to perform a full blown two-loop computation for these string backgrounds, it would nevertheless be interesting to probe the one-loop structure of the S-matrix and related questions. 
 
\section*{Acknowledgements}
It is a pleasure to thank M. Abbott, S. Penati, O. Ohlsson Sax, B. Stefanski and K. Zarembo for useful discussions and comments. The work of PS was supported by a joint INFN and
Milano-Bicocca postdoctoral grant and a short term scientific mission grant from COST, ECOST-STSM-MP1210-020115-051529. PS furthermore thanks Nordita, Stockholm, for hospitality during the duration of this grant. The work of LW was supported by the ERC Advanced grant No.290456 "Gauge theory - string theory duality".

\section*{Appendix}
\appendix
\section{Sunset integrals for massive propagator in $AdS_5 \times S^5$}
The full sunset contribution, before using any integral identities and going on-shell, is given by (here $I^{rstu}=I^{rstu}_{111}(p)$)
\bea
\label{eq:full-ads5-sun}
&& g^2\mathcal{A}_{sun} =
\frac{1}{4}\left(-4p_+^4+7p_+^3p_-+6p_+^2-22p_+^2p_-^2+12p_+p_-+3p_+p_-^3+6p_-^2-4p_-^4\right)I^{0000}
\\ \nn
&&
+\frac{1}{16}\left(-17p_+^3-8p_++201p_+^2p_--90p_+p_-^2+128p_-+54p_-^3\right)I^{0001}
\\ \nn
&&
+\frac{1}{16}\left(54p_+^3+128p_+-102p_+^2p_-+137p_+p_-^2-56p_-+35p_-^3\right)I^{0010}
\\ \nn
&&
+\frac{1}{16}\left(-5p_+^2+92p_+p_--80-54p_-^2\right)I^{0101}
+\frac{1}{16}\left(51p_+^2-48p_+p_-+160-37p_-^2\right)I^{1001}
\\ \nn
&&
+\frac{1}{8}\left(45p_+^2+12p_+^2p_-^2-82p_+p_-+56-11p_-^2\right)I^{1100}
\\ \nn
&&
+\frac{1}{16}\left(-54p_+^2+68p_+p_--80+3p_-^2\right)I^{1010}
+\frac{1}{16}\left(-73p_+^2+4p_+^3p_-+107p_+p_--8-66p_-^2\right)I^{0200}
\\ \nn
&&
+\frac{1}{16}\left(-66p_+^2+59p_+p_--8+20p_+p_-^3-57p_-^2\right)I^{2000}
+\frac{1}{4}\left(-9p_+-4p_+^2p_--19p_-\right)I^{1101}
\\ \nn
&&
+\frac{1}{4}\left(-77p_++4p_+p_-^2-3p_-\right)I^{1110}
+\frac{1}{16}\left(138p_+-55p_-\right) I^{2010}
+\frac{1}{16}\left(19p_--78p_+\right)I^{2001}
\\ \nn
&&
+\frac{1}{16}\left(138p_--107p_+\right) I^{0201}
+\frac{1}{16}\left(43p_+-2p_-\right)I^{0210}
+\frac{1}{8}\left(-91p_+-20p_+p_-^2+41p_-\right)I^{2100}
\\ \nn
&&
+\frac{1}{8}\left(41p_+-20p_+^2p_--27p_-\right)I^{1200}
+\frac{1}{4}\left(11p_++14p_--14p_-^3\right)I^{3000}
\\ \nn
&&
+\frac{1}{4}\left(-2p_+^3-2p_++11p_-\right)I^{0300}
-\frac{3}{4}\left(4p_+p_-+9\right)I^{1111}
+\left(11-3p_-^2\right)I^{2110}
\\ \nn
&&
+\frac{1}{16}\left(-88p_+p_--117\right)I^{2101}
+\frac{1}{16}\left(-88p_+p_--117\right)I^{1210}
+\frac{1}{4}\left(9-4p_+^2\right)I^{1201}
\\ \nn
&&
+\frac{1}{16}\left(67-20p_-^2\right)I^{2011}
+\frac{1}{16}\left(-4p_+^2-29\right)I^{0211}
+\frac{1}{2}\left(-4p_+p_--7\right)I^{2200}
-\frac{15}{8}I^{2020}
\\ \nn
&&
+\frac{1}{8}\left(-12p_+p_--47\right)I^{2002}
-\frac{15}{8}I^{0202}
+\frac{1}{4}\left(30p_-^2-1\right)I^{3100}
-\frac{15}{8}I^{3010}
\\ \nn
&&
+\frac{1}{16}\left(104p_-^2-69\right)I^{3001}
+\frac{1}{4}\left(2p_+^2-5\right)I^{1300}
+\frac{1}{16}\left(-8p_+^2-57\right)I^{0310}
-\frac{15}{8}I^{0301}
\\ \nn
&&
+\frac{9}{2}p_-I^{2111}
+\frac{13}{2}p_+I^{1211}
+8p_-I^{2210}
+8p_+I^{2201}
+3p_+I^{2102}
+p_-I^{1220}
\\ \nn
&&
-9p_-I^{3101}
+5p_+I^{1310}
+\frac{7}{2}p_+I^{2300}
-\frac{7}{2}p_-I^{3200}
-\frac{13}{2}p_-I^{3002}
+\frac{1}{2}p_+I^{0320}
\\ \nn
&&
-\frac{1}{2}I^{2211}
+5I^{2112}
+\frac{7}{2}I^{3201}
-\frac{7}{2}I^{2310}
+\frac{31}{4}I^{3102}
+\frac{7}{4}I^{1320}
+4I^{3003}\,.
%%%%%%%%
%=
%-\frac{16}{3}p_1^4I^{0000}
%-\frac{3}{2}(7+8p_1^2)[T_1^{00}]^2
\eea
The bubble tadpole diagrams give
\bea
\label{eq:full-ads5-bt}
\nn
&& 
g^2\mathcal{A}_{bt}=
4p_-^2B^{11}_1(0)T^{00}_1
-4p_-^2B^{00}_1(0)T^{11}_1
-4p_+p_-B^{11}_1(0)T^{00}
+4p_+p_-B^{00}_1(0)T^{11}_1
\\ 
&&
+4p_+^2B^{11}_1(0)T^{00}_1
-4p_+^2B^{00}_1(0)T^{11}_1
-4B^{22}_1(0)T^{00}_1
+4B^{11}_1(0)T^{11}_1\,.
%%%%%%%
%=16p_1^2[T^{00}]^2
\eea
From the six-vertex tadpoles we get
\bea
\label{eq:full-ads5-t6} 
\nn
&&g^2\mathcal{A}_{t_6}=
\frac{1}{2}(-2p_+^2+13p_+p_--3-2p_-^2)[T^{00}_1]^2
-2(p_+p_--6)T^{11}_1T^{00}_1
-\frac{5}{2}[T^{11}_1]^2\,.
%%%%%%%
%=-4p_1^2[T^{00}]^2+\frac{21}{2}[T^{00}]^2
\eea

\section{Decoupling of massless modes in $AdS_3\times S^3\times T^4$}
Here will collect the sunset integrals with massless virtual particles contributing to the massive mode dispersion relation. To keep the expression tractable we put $q=0$ for simplicity (although we have verified that the massless modes decouple for arbitrary $q$)
\bea 
\label{eq:ads3-massive-massless}
\nn
&&g^2\mathcal{A}_{sun}=
\frac18\left(-2p_-^3+9p_+p_-^2-p_+^2p_-\right)I_{1\mu\mu}^{0001}
+\frac18\left(-2p_+^3+17p_+^2p_--p_+p_-^2\right)I_{1\mu\mu}^{0010}
\\ \nn
&&
+\frac18\left(-2p_+^2-31p_+p_-+p_-^2\right)I_{1\mu\mu}^{0020}
+\frac18\left(p_+^2-7p_+p_--2p_-^2\right) I_{1\mu\mu}^{0002}
-p_+p_-I_{1\mu\mu}^{0011}
\\ \nn
&&
+\frac18\left(-4p_+^2-30p_+p_-+p_-^2\right)I_{1\mu\mu}^{1010}
+\frac18\left(p_+^2-14p_+p_--4p_-^2\right)I_{1\mu\mu}^{0101}
+\frac18\left(p_-^2-2p_+p_-\right)I_{1\mu\mu}^{1001}
\\ \nn
&&
+\frac18\left(p_+^2-8p_+^2p_-^2-2p_+p_-\right)I_{1\mu\mu}^{0110}
+\frac14\left(2p_--p_+\right)I_{1\mu\mu}^{0003}
+\frac14\left(2p_++7p_-\right)I_{1\mu\mu}^{0030}
\\ \nn
&&
+p_+I_{1\mu\mu}^{0012}
+p_-I_{1\mu\mu}^{0021}
+\frac18\left(18p_++27p_-\right)I_{1\mu\mu}^{1020}
+\frac18\left(3p_++18p_-\right)I_{1\mu\mu}^{0102}
\\ \nn
&&
+\frac18\left(p_++16p_+p_-^2+2p_-\right)I_{1\mu\mu}^{0120}
+\frac18\left(2p_+-8p_+^2p_-+p_-\right)I_{1\mu\mu}^{1002}
+\frac18\left(10p_++13p_-\right)I_{1\mu\mu}^{2010}
\\ \nn
&&
+\frac18\left(5p_++10p_-\right)I_{1\mu\mu}^{0201}
-\frac18p_+I_{1\mu\mu}^{0210}
-\frac18p_-I_{1\mu\mu}^{2001}
+\frac14\left(p_++8p_+p_-^2+p_-\right)I_{1\mu\mu}^{1110}
\\ \nn
&&
+\frac14\left(p_++p_-\right)I_{1\mu\mu}^{1101}
+p_+I_{1\mu\mu}^{0111}
+p_-I_{1\mu\mu}^{1011}
-\frac12I_{1\mu\mu}^{3010}
-\frac12I_{1\mu\mu}^{0301}
+\frac14\left(4p_+^2-1\right)I_{1\mu\mu}^{1003}
\\ \nn
&&
+\frac14\left(-4p_-^2-1\right)I_{1\mu\mu}^{0130}
-I_{1\mu\mu}^{1030}
-I_{1\mu\mu}^{0103}
-\frac{3}{2}I_{1\mu\mu}^{2020}
-\frac{3}{2}I_{1\mu\mu}^{0202}
+\frac18I_{1\mu\mu}^{2002}
+\frac18I_{1\mu\mu}^{0220}
-I_{1 \mu \mu}^{0022}
\\ \nn
&&
+\frac18\left(-16p_-^2-5\right)I_{1\mu\mu}^{1120}
+\frac18\left(8p_+^2-5\right)I_{1 \mu \mu}^{1102}
+\frac18\left(-8p_-^2-3\right)I_{1\mu\mu}^{2110}
+\frac18I_{1\mu\mu}^{2101}
\\ \nn
&&
+\frac18I_{1\mu\mu}^{1210}
-\frac{3}{8}I_{1\mu\mu}^{1201}
-I_{1\mu\mu}^{1012}
-I_{1\mu\mu}^{0121}
-I_{1\mu\mu}^{1111}
+\dots +\mathcal{O}(\mu)\,,
%%%%%%%%
%=
%-\mu^2p_+I_{1\mu\mu}^{2010}
%-\mu^2p_-I_{1\mu\mu}^{0201}
%+\frac14\left(2p_++7p_-\right)\mu^2p_+I_{1\mu\mu}^{1010}
%+\frac14\left(2p_--p_+\right)\mu^2p_-I_{1\mu\mu}^{0101}
%-\frac12\mu^2p_+I_{1\mu\mu}^{3000}
%+\frac18\left(6p_++13p_-\right)\mu^2p_+I_{1\mu\mu}^{2000}
%+\frac18\left(5p_++6p_-\right)\mu^2p_-I_{1\mu\mu}^{0200}
%-\frac12\mu^2p_-I_{1\mu\mu}^{0300}
%+\frac14\left(4p_+^2-1\right)\mu^2p_-I_{1\mu\mu}^{0001}
%+\frac14\left(-4p_-^2-1\right)\mu^2p_+I_{1\mu\mu}^{0010}
%+\frac18\left(2p_+^2-20-7p_-^2\right)\mu^2p_+I_{1\mu\mu}^{1000}
%+\frac18\left(p_+^2-12+2p_-^2\right)\mu^2p_-I_{1\mu\mu}^{0100}
%+\frac18\mu^2p_+I_{1\mu\mu}^{0100}
%+\frac18\mu^2p_-I_{1\mu\mu}^{1000}
%+\frac18\left(-p_++10p_-\right)\mu^2p_+I_{1\mu\mu}^{0000}
%+\frac18\left(2p_+-p_-\right)\mu^2p_-I_{1\mu\mu}^{0000}
%+\mu^2p_+I_{1\mu\mu}^{0100}
%+\mu^2p_-I_{1\mu\mu}^{1000}
%-\mu^2I_{1\mu\mu}^{0000}
%-\mu^2I_{1\mu\mu}^{0000}
%+\mu^2p_+I_{1\mu\mu}^{0001}
%+\mu^2p_-I_{1\mu\mu}^{0010}
%-\mu^2I_{1\mu\mu}^{0110}
%-\mu^2I_{1\mu\mu}^{1001}
%-\mu^2I_{1\mu\mu}^{0011}
%-\mu^2\frac54I_{1\mu\mu-1}^{0000}
%-\mu^2I_{1\mu\mu-2}^{0000}
\eea 
where the dots denote integrals with massive virtual particles only. At first glance it is not at all apparent that these expressions will cancel out. However, using the integral identities in section \ref{sec:disp-relation} one finds that the integrals indeed cancel among themselves demonstrating the decoupling of the massless modes from the massive dispersion relation.

\section{Sunset integrals for massless propagator in $AdS_3\times S^3\times T^4$ }
For general values of $q$ the sunset contribution equals
\bea
\label{eq:full-ads3-sun}
&& g^2\mathcal{A}_{sun}=
\frac14\mu(q\hat qp_+-q\hat qp_--8q^2\mu+4\mu )p_+p_-I_{\mu \mu \mu}^{1100}
+\frac14\mu\hat q(qp_+-2\hat q\mu)p_-^2I_{\mu\mu\mu}^{2000}
\\ \nn 
&&
-\frac14\mu\hat q(qp_-+2\hat q\mu)p_+^2I_{\mu\mu\mu}^{0200}
-\hat q^2p_-^3I_{\mu\mu\mu}^{2010}
+\frac{1}{4}q\hat q\mu(p_-^2-p_+p_-)I_{\mu\mu\mu}^{2100}
-\frac{1}{4}q\hat q\mu(p_+^2-p_+p_-)I_{\mu\mu\mu}^{1200}
\\ \nn 
&&
-\frac{1}{2}q(qp_++\hat q\mu)p_+p_-I_{\mu\mu\mu}^{2001}
+\frac12(-2\hat q^2p_++q\hat q\mu-q^2p_-)p_+p_-I_{\mu \mu \mu}^{0210} 
+\frac{1}{2}q\hat q\mu p_-^2I_{\mu\mu\mu}^{1110}
\\ \nn 
&&
-\frac{1}{2}q\hat q\mu p_+^2I_{\mu\mu\mu}^{1101}
+\frac{1}{2}q^2(p_+^2+p_-^2)I_{\mu\mu\mu}^{2002}
+\frac{1}{2}q^2p_-^2I_{\mu\mu\mu}^{1210}
+\frac{1}{2}q^2p_+^2I_{\mu\mu\mu}^{2101}
+\hat q^2p_-^2I_{\mu\mu\mu}^{2011}
\\ \nn 
&&
+\hat q^2p_+^2I_{\mu\mu\mu}^{0211}
+\hat q^2p_-^2I_{\mu\mu\mu}^{2110}
+\hat q^2p_+^2I_{\mu\mu\mu}^{0310}
+\frac{3}{2}q\hat q^3\mu(p_--p_+)p_+p_-I_{\mu\hat q\hat q}^{0000}
\\ \nn
&&
+\frac12\hat q^2(-\hat q^2p_+^3-p_+^2p_-+3q\hat q\mu p_+p_-)I_{\mu\hat q\hat q}^{1000}
+\frac12\hat q^2(-\hat q^2p_-^3-3q\hat q\mu p_+p_-+3(2q^2-1)p_+p_-^2)I_{\mu\hat q\hat q}^{0100}
\\ \nn
&&
+\frac12q\hat q\mu(p_+p_-+6\hat q^2)p_+p_-I_{\mu\hat q\hat q}^{0010}
+\frac12q\hat q\mu(-p_+p_--6\hat q^2)p_+p_-I_{\mu\hat q\hat q}^{0001}
-\frac14q^2p_+^2p_-^2I_{\mu\hat q\hat q}^{0200}
\\ \nn
&&
-\frac12q\hat q\mu p_+p_-^2I_{\mu\hat q\hat q}^{0020}
+\frac12q\hat q\mu p_+^2p_-I_{\mu\hat q\hat q}^{0002}
+q\hat q\mu(p_--p_+)p_+p_-I_{\mu\hat q\hat q}^{0011}
\\ \nn
&&
+\frac18(4\hat q^2p_+^2-8\hat q^4p_+p_-+q^2p_+^2p_-^2+12(\hat q^2-q^2\hat q^2+q^4)p_-^2)I_{\mu\hat q\hat q}^{1100}
\\ \nn
&&
+\frac12\hat q(-q\mu p_+^2p_-+2(1-3q^2)\hat qp_-^2-2\hat qp_+^2p_-^2)I_{\mu\hat q\hat q}^{0110}
+\frac18q(qp_-^2-4\hat q\mu p_+-4qp_+^2)p_-^2I_{\mu\hat q\hat q}^{1010}
\\ \nn
&&
+\frac12\hat q(q\mu p_-^2-2(1-3q^2)\hat qp_+)p_+I_{\mu\hat q\hat q}^{1001}
+\frac12q(\hat q\mu-qp_-)p_+^2p_-I_{\mu\hat q\hat q}^{0101}
+\frac{1}{4}q^2p_+^2p_-I_{\mu\hat q\hat q}^{2001}
\\ \nn
&&
+\frac{3}{4}q^2p_+p_-^2I_{\mu\hat q\hat q}^{0210}
-\frac{1}{8}q^2p_+p_-^2I_{\mu\hat q\hat q}^{2100}
+\frac18q^2(4p_--p_+)p_+p_-I_{\mu\hat q\hat q}^{1200}
-q\hat q\mu p_+p_-I_{\mu\hat q\hat q}^{0012}
+q\hat q\mu p_+p_-I_{\mu\hat q\hat q}^{0021}
\\ \nn
&&
+\frac12(q\hat q\mu p_++4\hat q^2p_+p_--p_-^2q^2)p_-I_{\mu\hat q\hat q}^{0120}
+\frac18(-4p_+^2q^2-4q\hat q\mu p_--8\hat q^2p_+p_-+q^2p_-^2)p_+I_{\mu\hat q\hat q}^{1002}
\\ \nn
&&
+\frac18(4q\hat q\mu p_++8q^2p_+^2+16p_+p_--19q^2p_+p_--2p_-^2q^2)p_-I_{\mu\hat q\hat q}^{1110}
+q(qp_--\hat q\mu)p_+p_-I_{\mu\hat q\hat q}^{0111}
\\ \nn
&&
+\frac18q(8qp_+p_-^2-4\hat q\mu p_+p_--3qp_+^2p_-)I_{\mu\hat q\hat q}^{1101}
+\frac14q(-qp_-^3+4qp_+^2p_-+4\hat q\mu p_+p_-)I_{\mu\hat q\hat q}^{1011}
\\ \nn
&&
+q^2p_-^2I_{\mu\hat q\hat q}^{0130}
-q^2p_+^2I_{\mu\hat q\hat q}^{1003}
-q^2p_+^2I_{\mu\hat q\hat q}^{1102}
-\frac{3}{4}q^2p_-^2I_{\mu\hat q\hat q}^{1111}
-q^2p_+^2I_{\mu\hat q\hat q}^{1111}
+q^2p_+p_-I_{\mu\hat q\hat q}^{1111}
\\ \nn
&&
+2q^2p_-^2I_{\mu\hat q\hat q}^{1120}
-\frac{5}{8}q^2p_-^2 I_{\mu\hat q\hat q}^{1210}
-\frac{1}{2}q^2p_+^2 I_{\mu\hat q\hat q}^{1210}
+\frac{3}{8}q^2p_+p_-I_{\mu\hat q\hat q}^{1210}
-\frac{1}{8}q^2p_-^2I_{\mu\hat q\hat q}^{2002}
-\frac{1}{2}q^2p_-^2I_{\mu\hat q\hat q}^{2101}
\\ \nn
&&
-\frac{1}{4}q^2p_+^2I_{\mu\hat q\hat q}^{2101}
+\frac{3}{8}q^2p_+p_-I_{\mu\hat q\hat q}^{2101}
+q^2p_-^2I_{\mu\hat q\hat q}^{2110}
-\frac{1}{4}q^2p_-^2I_{\mu\hat q\hat q}^{2200}
+\frac{1}{8}q^2p_+p_-I_{\mu\hat q\hat q}^{2200}
\\ \nn
&&
-p_-^2I_{\mu\hat q\hat q}^{0130}
+p_+^2I_{\mu\hat q\hat q}^{1003}
+p_+^2I_{\mu\hat q\hat q}^{1102}
-2p_-^2I_{\mu\hat q\hat q}^{1120}
-p_-^2I_{\mu\hat q\hat q}^{2110}\,,
%%%%%%%%
%=
%\mu^2(-\frac12q^2(p_+^2+p_-^2)+\mu^2(1-2q^2))[T_\mu^{00}]^2
%+\mu(finite)
%q=0:
%-\frac12p_+^3I_{\mu\hat q\hat q}^{1000}
%-\frac12p_-^3I_{\mu\hat q\hat q}^{0100}
%+(p_+^2+p_-^2)[T_m^{00}]^2
\eea
where $\mu$ is the IR-regulator mass for the massless modes. For bubble-tadpoles we find
\begin{align}
\nn
g^2\mathcal{A}_{bt}=&
\frac12\hat q^2p_+p_-B_\mu^{11}(0) T_\mu^{11}
-\frac23\hat q^2p_+p_-B_\mu^{11}(0)\left(T_{\hat q}^{11}-\hat q^2 T_{\hat q}^{00}\right)
\\ \nn 
&{}
+2q^2\left(p_+^2+p_-^2\right)\left(-\hat q^2B_{\hat q}^{00}(0)-\frac12B_{\hat q}^{11}(0)\right)T_\mu^{11}
+2q^2\hat q^4\left(p_+^2+p_-^2\right)B_{\hat q}^{00} (0)T_{\hat q}^{00}
\\ \nn 
&{}
-\frac32q^2\left(p_+^2+p_-^2\right)B_{\hat q}^{11}(0)T_{\hat q}^{11}
-\frac12\left(2-5q^2\right)\hat q^2\left(p_+^2+p_-^2\right)B_{\hat q}^{11}(0)T_{\hat q}^{00}
\\ 
&{}
+\left(1-3q^2\right)\hat q^2\left(p_+^2+p_-^2\right)B_{\hat q}^{00}(0)T_{\hat q}^{11}
\,,
\label{eq:full-ads3-bt}
%%%%%
%=-4\hat q^4p_1^2[T_{\hat q}^{00}]^2
\end{align}
which is naively IR-divergent before rewriting the bubble integrals using (\ref{eq:triv-identity}). For the six-vertex tadpoles we find
\begin{equation}
\label{eq:full-ads3-t6}
g^2\mathcal{A}_{t_6} = 
\frac12q^2\hat q^2(p_+-p_-)^2[T_{\hat q}^{00}]^2
+\hat q^2p_+p_-[T_{\hat q}^{00}]^2
+\frac14q^2(p_+-p_-)^2T_{\hat q}^{00}T_{\hat q}^{11}
-p_+p_-T_{\hat q}^{00}T_{\hat q}^{11}
\,.
%%%%%
%=3q^2\hat q^2p_1^2[T_{\hat q}^{00}]^2
\end{equation}

\bibliographystyle{JHEP} 

\bibliography{ads2ads3etc}

\end{document}

%% file: dgms.tex
%%
%% This file contains descriptions of feynman diagrams for use with the package feynMP
%% After running the main file, run "mpost diagrams" once to generate these
%%
%%
%% These are for propagators paper with Per.
%% The version in this folder edited 18 May 2011 to add unlabled ones for introduction.
%%

\newsavebox{\feynmanrules}
\sbox{\feynmanrules}{
\begin{fmffile}{diagrams} % I can't seem to make this work using any path but the same one as the document

%%%%%%%%%%%%%%%%
%%  SETTINGS

\fmfset{thin}{0.6pt}  % was 0.7 until v24
%\fmfset{wiggly_len}{5mm}
\fmfset{dash_len}{4pt}
\fmfset{dot_size}{1thick}
\fmfset{arrow_len}{6pt} % you can't use em here, mpost doesn't know what it will be.
%\fmfset{curly_len}{2.5mm}
%\setlength{\unitlength}{1em} % default is =1pt, maybe that's sensible. 72pt = 1in

%%%%%%%%%%%%%%%%%%%%%%%%%%% AdS3 - Scatterings %%%%%%%%%%%%%%%%%%%%%%%%%%%%%%%

%%%% SUNSET

\begin{fmfgraph}(80,40)
\fmfkeep{sunset}
\fmfleft{i}
\fmfright{o}
\fmf{plain,tension=5}{i,v1}
\fmf{plain,tension=5}{v2,o}
\fmf{plain,left,tension=0.4}{v1,v2,v1}
\fmf{plain}{v1,v2}
\fmfdot{v1,v2}
\end{fmfgraph}

%%% DOUBLE BUBBLE
\begin{fmfgraph*}(120,75)
\fmfkeep{doubblebubble}
    \fmfleft{i1,i2,i3}
    \fmfright{o1,o2,o3}
    \fmf{plain}{i1,v1,o1}
    \fmffreeze
    \fmf{phantom}{i2,v2,o2}             
    \fmf{phantom}{i3,v3,o3}
    \fmf{plain,left}{v1,v2,v1}
    \fmf{plain,left}{v2,v3,v2}
    \fmfdot{v1,v2}
\end{fmfgraph*}

\begin{fmfgraph*}(72,25)
\fmfkeep{single}
\fmfleft{in,p1}
\fmfright{out,p2}
\fmfdot{c}
\fmf{dashes_arrow,label=\small{new}}{in,c}
\fmf{dashes_arrow}{c,out}
\fmf{plain_arrow,right, tension=0.8, label=\small{lables}}{c,c}
\fmf{phantom, tension=0.2}{p1,p2}
\end{fmfgraph*}

%%% BUBBLE

\begin{fmfgraph*}(80,40)
\fmfkeep{schannel}
\fmfleft{i1,i2}
\fmfright{o1,o2}
\fmf{plain}{i1,v1}
\fmf{plain}{i2,v1}
\fmf{plain,left=0.5,tension=0.4}{v1,v2}
\fmf{plain,right=0.5,tension=0.4}{v1,v2}
\fmf{plain}{v2,o1}
\fmf{plain}{v2,o2}
\fmfdot{v1,v2}
\end{fmfgraph*}

\begin{fmfgraph*}(80,40)
\fmfkeep{tchannel}
\fmfleft{i1,i2}
\fmfright{o1,o2}
\fmf{plain}{i1,v1,o1}
\fmf{plain}{i2,v2,o2}
\fmf{plain,left=0.5,tension=0.4}{v1,v2}
\fmf{plain,right=0.5,tension=0.4}{v1,v2}
%\fmf{plain}{v2,o1}/
%\fmf{plain}{v2,o2}
\fmfdot{v1,v2}
\end{fmfgraph*}

\begin{fmfgraph*}(80,40)
\fmfkeep{uchannel}
\fmfleft{i1,i2}
\fmfright{o1,o2}
\fmf{plain}{i1,v1}
\fmf{phantom}{v1,o1} % Invisible rubber band
\fmf{plain}{i2,v2}
\fmf{phantom}{v2,o2} % also invisible rubber band
\fmf{plain,left=0.5,tension=0.4}{v1,v2}
\fmf{plain,right=0.5,tension=0.4}{v1,v2}
% These are visible, but have no tension.
\fmf{plain,tension=0}{v1,o2}
\fmf{plain,tension=0}{v2,o1}
\fmfdot{v1,v2}
\end{fmfgraph*}

\begin{fmfgraph*}(80,40)
\fmfkeep{tadpolesix}
\fmfbottom{i1,o1}
\fmftop{i2,o2}
\fmf{plain}{i1,v1,o1}
\fmf{plain}{i2,v1,o2}
\fmf{plain,right=90,tension=0.8}{v1,v1}
%\fmf{plain,label=$(s)$}{v1}
%\fmf{fermion,tension=0}{v1,v2}
\fmfdot{v1}
\end{fmfgraph*}

%%%%%%%%%%%%%%% Two pouint functions

\begin{fmfgraph*}(100,36)
\fmfkeep{bubble}
\fmfleft{in}
\fmfright{out}
\fmfdot{v1}
\fmfdot{v2}
\fmf{plain}{in,v1}
\fmf{plain}{v2,out}
\fmf{plain,left,tension=0.6}{v1,v2}
\fmf{plain,right,tension=0.6}{v1,v2}
\end{fmfgraph*}

%%% TADPOLE

%\begin{fmfgraph*}(72,25)
\begin{fmfgraph*}(100,36)
\fmfkeep{tadpole}
\fmfset{dash_len}{6pt} % this seems to be a local change
\fmfleft{in,p1}
\fmfright{out,p2}
\fmfdot{c}
\fmf{plain}{in,c}
\fmf{plain}{c,out}
\fmf{plain,right, tension=0.8}{c,c}
\fmf{phantom, tension=0.2}{p1,p2}
\end{fmfgraph*}

\begin{fmfgraph*}(100,36)
\fmfkeep{doubletadpole}
\fmfpen{thin}
\fmfleft{i}
\fmfright{o}
\fmf{plain}{i,v,v,o}
\fmf{plain,left=90}{v,v}
\fmfdot{v}
\end{fmfgraph*}

\end{fmffile}

}